\documentclass[sigconf,screen]{acmart}
\usepackage{graphicx}
\usepackage{subfiles}
\usepackage{array}
\usepackage{amsmath}
\usepackage{algorithm}
\usepackage{multirow}
\usepackage{subfigure}
\usepackage{caption}
\usepackage{flushend}
\usepackage{balance}
\usepackage{diagbox}
\usepackage{ulem}
\usepackage{algorithmicx}
\usepackage{colortbl}
 \usepackage{enumitem}
 \usepackage{balance}
\usepackage{CJKutf8}
\usepackage{supertabular,booktabs}
\usepackage{fancyhdr}

\usepackage{balance}

\usepackage{listings}
\usepackage{color}

\definecolor{dkgreen}{rgb}{0,0.6,0}
\definecolor{gray}{rgb}{0.5,0.5,0.5}
\definecolor{mauve}{rgb}{0.58,0,0.82}

\usepackage{soul}
\usepackage{makecell}
\usepackage{tcolorbox}
\usepackage[noend]{algpseudocode}
\usepackage{float}
\usepackage{boxedminipage}
\usepackage{fancybox}
\usepackage{xcolor}
\usepackage{adjustbox}
\usepackage{graphicx}

\definecolor{codegreen}{rgb}{0,0.6,0}
\definecolor{codegray}{rgb}{0.5,0.5,0.5}
\definecolor{codepurple}{rgb}{0.58,0,0.82}
\definecolor{backcolour}{rgb}{0.95,0.95,0.92}

\lstdefinestyle{mystyle}{
  language=Java,
  aboveskip=3mm,
  showstringspaces=false,
  columns=flexible,
  numbers=none,
  backgroundcolor=\color{backcolour},
  commentstyle=\color{codegreen},
 keywordstyle=\color{magenta},
    numberstyle=\tiny\color{codegray},
    stringstyle=\color{codepurple},
    basicstyle=\small\ttfamily,
    breakatwhitespace=false,         
    breaklines=false,                 
    captionpos=b,                    
    keepspaces=false,                 
    numbersep=5pt,                  
    showspaces=false,                
    showstringspaces=false,
    showtabs=false,                  
    tabsize=2,
    escapeinside=``
}
\newcommand{\pcsd}{\texttt{PCSD}}
\newcommand{\tlc}{\texttt{TLC}}
\newcommand{\funcom}{\texttt{Funcom}}
\newcommand{\csn}{\texttt{CSN}}
\newcommand{\codenn}{\texttt{CSN}}

\newcommand{\tool}{CAT}

\newcommand{\RQone}{data preprocessing errors}
\newcommand{\RQtwo}{low-quality comments}

\usepackage{xcolor}
\usepackage{calc}

\definecolor{color1}{rgb}{0.22,0.45,0.70}  % light blue
\definecolor{color2}{rgb}{0.45,0.45,0.45}  % dark grey

\usepackage{algorithm}
\usepackage{bm}
\usepackage{amsthm}
\usepackage{amsmath}
\usepackage{mathrsfs}
\usepackage{subfigure}
\usepackage{flushend}

\setcopyright{rightsretained}
\acmPrice{}
\acmDOI{10.1145/3540250.3549145}
\acmYear{2022}
\copyrightyear{2022}
\acmSubmissionID{fse22main-p766-p}
\acmISBN{978-1-4503-9413-0/22/11}
\acmConference[ESEC/FSE '22]{Proceedings of the 30th ACM Joint European Software Engineering Conference and Symposium on the Foundations of Software Engineering}{November 14--18, 2022}{Singapore, Singapore}
\acmBooktitle{Proceedings of the 30th ACM Joint European Software Engineering Conference and Symposium on the Foundations of Software Engineering (ESEC/FSE '22), November 14--18, 2022, Singapore, Singapore}

\begin{document}

\title{Are We Building on the Rock? On the Importance of Data Preprocessing for Code Summarization}

\author{Lin Shi}
\authornote{Also With Laboratory for Internet Software Technologies, Institute of Software, CAS}
\authornote{Also With University of Chinese Academy of Sciences}
\authornote{Both authors contributed equally to this research}
\email{shilin@iscas.ac.cn}
\affiliation{%
\institution{Institute of Software, Chinese Academy of Sciences}
\city{Beijing}
\country{China}
}

\author{Fangwen Mu}
\authornotemark[1]
\authornotemark[2]
\authornotemark[3]
\email{fangwen2020@iscas.ac.cn}
\affiliation{%
\institution{Institute of Software, Chinese Academy of Sciences}
\city{Beijing}
\country{China}
}

\author{Xiao Chen}
\authornotemark[1]
\authornotemark[2]
\email{chenxiao2021@iscas.ac.cn}
\affiliation{%
\institution{Institute of Software, Chinese Academy of Sciences}
\city{Beijing}
\country{China}
}

\author{Song Wang}
\email{wangsong@eecs.yorku.ca}
\affiliation{%
  \institution{Lassonde School of Engineering, York University}
  \city{Toronto}
  \country{Canada}
}

\author{Junjie Wang}
\authornotemark[1]
\authornotemark[2]
\email{junjie@iscas.ac.cn}
\affiliation{%
\institution{Institute of Software, Chinese Academy of Sciences}
\city{Beijing}
\country{China}
}

\author{Ye Yang}
\email{yangye@gmail.com}
\affiliation{%
  \institution{School of Systems and Enterprises, Stevens Institute of Technology}
  \city{Hoboken, NJ}
  \country{USA}
}

\author{Ge Li}
\email{lige@pku.edu.cn}
\affiliation{%
  \institution{Key Lab of High Confidence Software Technology, Peking University}
  \city{Beijing}
  \country{China}
}

\author{Xin Xia}
\email{xin.xia@acm.org}
\affiliation{%
  \institution{Software Engineering Application Technology Lab, Huawei}
%   \city{Beijing}
  \country{China}
}

\author{Qing Wang}
\authornotemark[1]
\authornotemark[2]
\authornote{Also With Science \& Technology on Integrated Information System Laboratory, Institute of Software, CAS}
\authornote{Corresponding author}
\email{wq@iscas.ac.cn}
\affiliation{
\institution{Institute of Software, Chinese Academy of Sciences}
\city{Beijing}
\country{China}
}

% \author{Lin Shi$^{1,3}$,  Fangwen Mu$^{1,3}$, Xiao Chen$^{1,3}$,  Song Wang$^{4}$,
%  Junjie Wang$^{1,3}$, \\ Ye Yang$^{5}$, Ge Li$^{6}$,  Xin Xia$^{7}$, Qing Wang$^{1,2,3}$}

% \affiliation{ $^1$Laboratory for Internet Software Technologies,}
% \affiliation{$^2$ State Key Laboratory of Computer Science, \\Institute of Software Chinese Academy of Sciences
% \city{Beijing}
% \country{China}}
% \affiliation{$^3$ University of Chinese Academy of Sciences
% \city{Beijing}
% \country{China}}
% \affiliation{$^4$ York University, Lassonde School of Engineering
% \country{Canada}}
% \affiliation{$^5$ School of Systems and Enterprises, Stevens Institute of Technology
% \city{Hoboken, NJ}
% \country{USA}}
% \affiliation{$^6$ Peking University, Key Lab of High Confidence Software Technology
% \city{Beijing}
% \country{China}}
% \affiliation{$^7$ Huawei, Software Engineering Application Technology Lab
% \country{China}
% \{shilin,fangwen2020,chenxiao2021,junjie,wq\}@iscas.ac.cn,\\
% wangsong@eecs.yorku.ca,
% yangye@gmail.com, 
% lige@pku.edu.cn,
% xin.xia@acm.org
% }

% \authornote{Corresponding author. }

% \renewcommand{\shortauthors}{Lin Shi et al.}
% \renewcommand{\authors}{Lin Shi, Fangwen Mu, Xiao Chen, Song Wang, Junjie Wang, Ye Yang, Ge Li, Xin Xia, Qing Wang}

\begin{abstract}
\vspace{0.1cm}
Code summarization, the task of generating useful comments given the code, has long been of interest. Most of the existing code summarization models are trained and validated on widely-used code comment benchmark datasets. However, little is known about the quality of the benchmark datasets built from real-world projects. Are the benchmark datasets as good as expected?
To bridge the gap, we conduct a systematic research to assess and improve the quality of four benchmark datasets widely used for code summarization tasks. First, we propose an automated code-comment cleaning tool that can accurately detect noisy data caused by inappropriate data preprocessing operations from existing benchmark datasets. Then, we apply the tool to further assess the data quality of the four benchmark datasets, based on the detected noises. Finally, we conduct comparative experiments to investigate the impact of noisy data on the performance of code summarization models. The results show that these data preprocessing noises widely exist in all four benchmark datasets, and removing these noisy data leads to a significant improvement on the performance of code summarization. We believe that the findings and insights will enable a better understanding of data quality in code summarization tasks, and pave the way for relevant research and practice.
\end{abstract}




\maketitle

\vspace{0.25cm}
\section{Introduction}
\vspace{0.05cm}

\begin{figure*}[t]
\centering
%\vspace{-0.3cm}
\includegraphics[width=\textwidth, height=2.8cm]{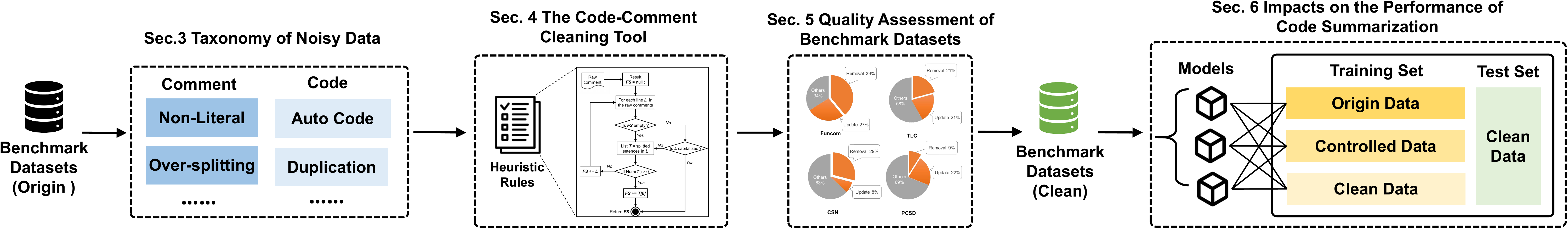}
%\vspace{-2.8cm}
\caption{Overview of our research methodology.} %Ye: should we include the taxonomy in the overview figure? Also, is it a good idea to have Figure 2 in Intro section? Or move it to later. The fact that Section 3-6 are individual components, and Fig. 2 in Intro section, makes the fluent of the paper somewhat broken.}
\label{fig:overview}
\vspace{-0.4cm}
\end{figure*}

Code summarization concerns the production of a natural-language description of source code that facilitates
software development and maintenance by enabling developers to comprehend, ideate, and document code effectively.
Learning-based models have been widely leveraged for the advantages in semantic modeling and understanding of languages. 
Similar to many other learning tasks, code summarization models require large-scale and high-quality training datasets. To that end, multiple benchmark datasets for code summarization tasks have been constructed from real-world project repositories, \textit{e.g.,} GitHub, and are popularly used {in many code summarization studies}. For example, {\funcom}~\cite{leclair2019neural} was released with over 2.1M code-comment pairs from over 29K Java projects in the Sourcerer repository. Many code summarization models, such as Re$^{2}$Com~\cite{wei2020retrieve}, DeepSumm~\cite{DBLP:journals/corr/abs-2004-00998}, and EditSum~\cite{DBLP:conf/kbse/LiL000J21}, are trained and evaluated to be relatively effective on it. 
%These models are evaluated to be relatively effective on generating code comments. 
Similar popular datasets include {\tlc}~\cite{hu2018summarizing}, {\codenn}~\cite{husain2019codesearchnet}, and {\pcsd} ~\cite{wan2018improving}.

Although the benchmark datasets are expected to be of good quality, 
%the comments are supposed to follow commenting conventions, and the code are supposed to be valid and coherent to comments, 
%the real-world nature make these large-volume benchmark datasets inevitable to comprise noises. 
noise is inevitable due to the differences in coding conventions and assumptions employed in modern programming languages and IDEs, as well as ad hoc nature of development processes and practices~\cite{DBLP:conf/scam/RaniASBN21}.
For example, source code in GitHub is contributed by developers all around the world, thus their comments are likely to contain multiple natural languages that can lead to increases in complexity regarding the understanding and maintenance of source code. 
%Actually, many other researchers have pointed out existing benchmark datasets still have noisy data, 
Existing studies also have confirmed the existence of many different types of noise in various benchmark datasets, such as auto-generated code~\cite{DBLP:conf/acl/HasanMIMHHAIS21}, ``TODO'' comments~\cite{DBLP:journals/tosem/ChenXHLL21}, and incomplete comments~\cite{DBLP:journals/corr/abs-2107-07112}, despite their data cleaning efforts. 
%only a limit number of specific types of noise are handled in existing studies.
{Particularly, Steidl \textit{et al.}~\cite{DBLP:conf/iwpc/SteidlHJ13} analyzed five open source projects, and reported that nearly one third of the comments do not promote system understanding.}
To investigate the aforementioned concerns of data quality for code summarization, we conduct a systematic study to assess and improve the quality of four widely-used benchmark datasets, \textit{i.e.}, {\funcom}, {\tlc}, {\codenn}, and {\pcsd}. The research methodology overview consists of four main steps, as illustrated in Figure~\ref{fig:overview}. 
%We choose the four widely used benchmark datasets, i.e., {\pcsd} ~\cite{wan2018improving}, {\tlc}~\cite{hu2018summarizing}, {\funcom}~\cite{leclair2019neural}, and {\codenn}~\cite{iyer2016summarizing}, as our studied benchmark datasets.
%\rev{First, we propose an automated code-summary cleaning tool that can precisely detect noisy data from existing benchmark datasets. Basically, it encompasses two subsequent cleaner: a rule-based syntactic cleaner based on syntax features and a model-based semantic cleaner trained on an external dataset.}
First, we propose a taxonomy of 12 different types of data noises due to inappropriate or insufficient data preprocessing in code summarization, 
derived from observations on the selected four benchmark datasets. %The taxonomy contains 12 different types of noises. 
Second, we build a rule-based cleaning tool, named {\tool} (Code-comment cleAning Tool), for {automatically scanning and} detecting {the occurrences and distribution of data} noises for a given dataset, based on the proposed taxonomy. 
The manual {validation} %inspection of the cleaning 
results show that the tool can accurately detect noisy data.
%Ye: "precisely" sounds vague and might be over-claiming. maybe replace with "accurately" or provide an accuracy metric.
%Second, 
{Third,} 
we conduct an evaluation study to assess the data quality of the four widely-used benchmark datasets. The results show that noisy data extensively exist in the four benchmark datasets (ranging from 31\% to 66\%).
%Third, 
{Finally,} we investigate the impacts of noises on three typical code summarization models (\textit{i.e.} NNGen~\cite{liu2018neural}, NCS~\cite{ahmad2020transformer}, and Rencos~\cite{zhang2020retrieval}) by comparing their performance trained on the same datasets before and after data cleaning. 
The above four steps will be elaborated in later sections, \textit{i.e.}, sec. 3 to sec. 6, respectively. 
The results show that, removing noisy data have a positive influence on model summarization ability.
%\song{negative influence? since removing noise boosts the performance}
Training three models with the filtered datasets improves the BLEU-4 by 27\%, 21\%, and 24\%, respectively.
%ROUGE by 19\%, 11\%, and 16\%, METEOR by 19\%, 7\%, and 16\%, CIDEr by 46\%, 19\%, and 33\%, 
%CIDEr, XXX 20.70\% BLEU-4 and 19.43\% CIDEr, 25.88\% BLEU-4 and 32.58\% CIDEr, respectively.
The major contributions of this paper are as follows.
\begin{itemize}[leftmargin=4mm]
    \item To the best of our knowledge, it is the first to systematically study the {patterns and} impact of the noises in various code summarization datasets. 
   \item We develop an automated data cleaning tool, named {\tool}, for code summarization datasets, which can help distill high-quality code-comment data.
    \item We perform a comprehensive assessment on data quality of benchmark datasests, which provides practical insights for future code summarization research.  
    \item %We conduct a comparative analysis of model performance trained on the origin and distilled clean benchmark datasets, which demonstrates significant model performance improvement.
    We conduct a comparative analysis on the performance of code summarization models trained on the origin and distilled benchmark datasets, our results demonstrate that removing noises yields significant model performance improvement.
     \item We release {\tool} and the distilled benchmark datasets~\cite{website} to the general public, in order to facilitate the replication of our study and its extensive application in other contexts. 
\end{itemize}

In the remainder of the paper, Section~\ref{sec:pre} illustrates the preliminaries. Section~\ref{sec:tox} introduces the taxonomy of noisy data. Section~\ref{sec:cleaningtool} presents the code-comment cleaning tool. Section~\ref{sec:qassessment} demonstrates the quality assessment of benchmark datasets. Section~\ref{sec:modelperformance} shows the impact of noisy data on the performance of code summarization. Section~\ref{sec:threats} discusses results and threats to validity. Section~\ref{sec:relatedwork} surveys the related work to our study.  Section~\ref{sec:conclusion} concludes this paper.

% For RQ1, we evaluate the syntactic noises in the provided data from summary level and code level. After discarding the syntactically noisy data, we obtain the SYN-clean datasets. 

% For RQ2, we evaluate the semantic noises and the quality of the provided summaries in the SYN-clean datasets by three means: 
% (1) We conduct a model-based evaluation to analyze the concentration of actual summaries. The model can predict whether a given sentence is a code summary or not.  
% (2) We conduct a human evaluation to analyze the accuracy, conciseness, and adequacy on the randomly sampled summaries. 
% (3) We conduct a metric-based evaluation to analyze the readability and coherence of the provided summaries.

% For RQ3, we investigate the performance variation of the state-of-the-art code summarization approaches after removing both syntactically and semantically noisy summaries in the training sets. 

\section{Preliminaries}
\label{sec:pre}

This section briefly introduces the literature of code summarization, as well as four widely-used benchmark datasets.

\subsection{Code Summarization}
\label{sec:codeSum}
%\ye{See comment.}
%Ye: This subsection provides background on different code summarization models, based on the underlying techniques used, e.g., IR, NMT, and hybrid. However, it is not clear how do these different techniques differs in terms of their assumptions for the input data, as well as the internal processing. It would be nice to provide some kind of discussion to link these pieces together. 
%\jie{In this para, how about we  explicitly state that there are three categories of code summarization models and organize the content aligned with the three categories,  which can be better mapped to the content of sec 5.1.2. In sec 5.1.2, we can say we choose the state-of-the-art model from each categories. }\lin{@wangsong}\song{updated}

{Code summarization~\cite{moreno2013automatic,sridhara2010towards} aims at generating a comment for a given block of source code that can help developers better understand and maintain source code.} 
%concerns the production of short natural language descriptions of source code that can help developers better understand and maintain source code. 
The essential task is to translate the code written in programming languages into {comments} written in natural languages. Meanwhile, {comments} may describe not only the functions, but also the design intents, program logic, and functionalities of programs behind the source code. 
%In the early stage of automatic code summarization, manually-craft templates were leveraged to generate comments automatically~\cite{moreno2013automatic,sridhara2010towards}. For example, Moreno et al.~\cite{moreno2013automatic} used predefined heuristic rules to capture information from the source code and further used them to generate summarization. As these template based approaches require expert domain knowledge, along this line, 
The existing code summarization models can be categorized into three different types based on the techniques used, \textit{i.e.}, Information Retrieval (IR) based approaches~\cite{eddy2013evaluating,haiduc2010supporting,wong2015clocom}, Neural Machine Translation (NMT) based approaches~\cite{ye2020leveraging,wan2018improving,wei2019code,leclair2020improved,haque2020improved,chen2018neural,cai2020tag,alon2018code2seq,ahmad2020transformer,iyer2016summarizing,wei2020retrieve}, and hybrid approaches~\cite{yuchao2021yet,hu2018deep,hu2020deep,leclair2019neural} that combine IR and NMT techniques. 

Specifically, IR-based code summarization models use IR techniques to extract keywords from the
source code and compose them into term-based summarization for a given code snippet~\cite{eddy2013evaluating,haiduc2010supporting,wong2015clocom}. For example, Edmund \textit{et al.}~\cite{wong2015clocom} generated code summarization for a given code snippet by retrieving the replicated code samples from the corpus with clone detection techniques. 
Recently, with the booming of deep learning techniques, many NMT based code summarization approaches have been proposed, which train the neural models from a large-scale code-comment corpus to automatically generate summaries~\cite{ye2020leveraging,wan2018improving,wei2019code,leclair2020improved,haque2020improved,chen2018neural,cai2020tag,alon2018code2seq,ahmad2020transformer,iyer2016summarizing,hu2018deep,wei2020retrieve}. For example, Iyer \textit{et al.}~\cite{iyer2016summarizing} treated the code summarization task as an end-to-end translation problem and first introduced NMT into code comment generation. The hybrid approaches~\cite{zhang2020retrieval,yuchao2021yet,hu2020deep,leclair2019neural} leverage the advantages of IR and NMT techniques for improving code summarization. For example, Zhang \textit{et al.}~\cite{zhang2020retrieval} first retrieved top similar code in the training data for a given piece of code and then input them into an NMT model for summarization generation. 

%Information Retrieval based approaches.

%Neural Machine Translation (NMT) based approaches.

%Hybrid approaches~\cite{yuchao2021yet,hu2018deep,hu2020deep,leclair2019neural}.

\subsection{Benchmark Datasets}
\begin{table}[b]
\vspace{-0.4cm}
\caption{Benchmark Datasets Information}
\vspace{-0.2cm}
\label{table:dataset_intro}
\resizebox{\columnwidth}{!}{
\begin{tabular}{|c|c|c|c|c|}
\hline
Name              & {\funcom}                              & {\tlc}                                 & {\codenn}                                 & {\pcsd}                                \\ \hline
Year              & 2019                                & 2018                                & 2019                                & 2017                                \\ \hline
Source            & Sourcerer                           & Github                              & Github                              & Github                              \\ \hline
Download          & \cite{dataset_Funcom}                            & \cite{dataset_TLC}  & \cite{dataset_CSN}  & \cite{dataset_PCSD_clean}  \\ \hline
Language          & Java                                & Java                                & Java                                & Python                              \\ \hline
\#Pairs           & 2,149,121                           & 87,136                              & 496,688                             & 105,540                             \\ \hline
Train/Val/Test    & 9/0.5/0.5 by project                      & 8/1/1 by function                         & 8/1/1 by project                         & 6/2/2 by function                         \\ \hline
Trained-on Models &\makecell{\cite{leclair2019neural,DBLP:conf/kbse/LiL000J21,DBLP:conf/kbse/GrosSDY20}\\\cite{haque2020improved,leclair2020improved,wei2020retrieve}\\\cite{DBLP:conf/emnlp/Shi0D0HZS21,DBLP:conf/iwpc/BansalHM21,DBLP:conf/wcre/HaqueBWM21}\\\cite{DBLP:journals/corr/abs-2106-08415,DBLP:conf/icsm/LeClairBM21,DBLP:journals/corr/abs-2004-00998}}  
&\makecell{\cite{hu2018summarizing,DBLP:journals/nca/ZhouYF21,wei2019code}\\\cite{DBLP:conf/emnlp/Shi0D0HZS21,ahmad2020transformer,DBLP:journals/corr/abs-2108-00213}\\\cite{DBLP:journals/corr/abs-2104-09340,DBLP:journals/corr/abs-2111-08874,DBLP:journals/corr/abs-2107-07112}\\\cite{DBLP:conf/internetware/ZhangZ0C20,zhang2020retrieval}  }
&\makecell{\cite{husain2019codesearchnet,DBLP:conf/icst/PourL0H21,DBLP:journals/corr/abs-2108-01585}\\\cite{DBLP:journals/corr/abs-2107-07112,DBLP:conf/iwpc/ShahbaziSF21,DBLP:journals/corr/abs-2107-01933}\\\cite{DBLP:conf/emnlp/FengGTDFGS0LJZ20,DBLP:conf/emnlp/0034WJH21,DBLP:journals/corr/abs-2102-04664}\\\cite{DBLP:conf/iclr/GuoRLFT0ZDSFTDC21,DBLP:conf/iwpc/LinOZCLW21,DBLP:conf/icse/MastropaoloSCNP21}}  
&\makecell{\cite{wan2018improving,DBLP:journals/corr/abs-2104-09340,DBLP:conf/acl/WuZZ21}\\\cite{DBLP:conf/kbse/GrosSDY20,wei2019code,ahmad2020transformer}\\\cite{DBLP:journals/corr/abs-2108-00213,DBLP:journals/corr/abs-2104-09340,DBLP:conf/acl/WuZZ21}\\\cite{DBLP:journals/corr/abs-2111-08874,DBLP:journals/tse/WangZSWZWYX22,zhang2020retrieval} } \\ \hline
\end{tabular}
}
\end{table}
\label{sec:bench}
%\ye{See comment.}
%Ye: Suggest to move this subsection to next Section. It logically fits better there. 
{As introduced earlier, this study} conducts various experiments on four widely-used code summarization datasets, including {\funcom}~\cite{leclair2019neural}, {\tlc}~\cite{hu2018summarizing}, {\codenn}~\cite{husain2019codesearchnet}, and {\pcsd}~\cite{wan2018improving}. 
{The data format of these datasets is primarily represented using \textbf{code-comment pairs}, where the code data is at the granularity of \textbf{method-level}. Each dataset applies its own operations when extracting and preprocessing the raw data.}
%\ye{Some common characteristics of these datasets are: (1) data format is primarily represented using \textbf{code-comment pairs}; (2) data is extracted and measured at the granularity of \textbf{method-level}; and (3) each dataset applies certain assumptions when extracting and pre-processing the raw data.} 
Table~\ref{table:dataset_intro} {summarizes the information of descriptive metadata and associated studies where each dataset has been employed in existing literature}. 

{More specifically, }
\textbf{{\funcom}} is a collection of 2.1M code-comment pairs from 29K projects. For each method, it extracted its Javadoc comment and treated the first sentence in the Javadoc of each method as its summary. 
\textbf{{\tlc}} has 87K code-comment pairs collected from more than 9K open-source Java projects created from 2015 to 2016 with at least 20 stars. It extracted the Java methods and their corresponding Javadoc comments. These comments are considered as code summaries.
\textbf{{\csn}} %is code summarization dataset collected from Stack Overflow (SO), it focuses on SO posts about C\# and SQL. For a collected post, its title is considered as summary for the code
%snippets from the accepted answers. It has 145K code summary pairs for C\# and 41,340 code summary pairs for SQL. 
contains about 2M method and comment pairs mined from publicly available open-source non-fork GitHub repositories spanning six programming languages, \textit{i.e.}, Go, Java, JavaScript, PHP, Python, and Ruby. %\lin{In this study, we use Java XXX @fangwen}
{In this study, we conduct the experiments on the Java portion of the {\csn} dataset.} 
\textbf{{\pcsd}} contains 105K pairs of Python functions and their comments from open source repositories in GitHub. Specifically, it uses docstrings (\textit{i.e.}, the string literals that appear right after the definition of functions) as summaries for Python functions. 
%It contains 108K code-comment pairs.  
%It also contains automatically generated query-like natural
%language for 2 million functions, obtained from mechanically scraping and preprocessing associated function documentation. 

%table: name, year, source, url,language, train/val/test, usage

\section{The Taxonomy of Noisy Data}
\label{sec:tox}
%First, we introduce how we build the noisy data taxonomy. {Next, we introduce categories of comment-related and code-related noisy data caused by inappropriate data preprocessing operations, as well as their examples.} 

{An essential and effective starting point is a systematic and robust categorization of data noises. This section presents details on how the noisy data taxonomy is built, and the descriptions and examples for every 12 categories.}

\subsection{Taxonomy Construction}
\label{sec:3.1}
%To develop a taxonomy of noisy data for code summarization tasks, 
We employ an \textit{open card sort} \cite{DBLP:journals/es/RuggM05} process by involving nine participants. 
{Participants include two PhD students, four master students, and three senior researchers. 
All of them have done either intensive research work with software development or have been actively contributing to open-source projects.}
The sorting process is conducted in multiple rounds. 
{For each round, we randomly sample 160 code-comment pairs without replacement from the four benchmark datasets (40 pairs for each).} 
In the first round, 
all participants label the same sampled data, %\song{this is the first time mentioning sample data, should we give some info about the sample data early?}
% (a total of XXX pairs) \lin{@fangwen}
with an intensive discussion session to achieve conceptual coherence about noisy categories. The average Cohen’s Kappa is 0.86, which indicates substantial agreement. 
Then, a shared pool of categories is utilized and carefully maintained, and
each participant could select existing categories from and/or add new category names into the shared pool. The sorting process ends when there is no new category added for two consecutive rounds. In total, we conducted 10 rounds and labeled 1,600 pairs of source code and the corresponding comments (400 pairs for each of the four benchmark datasets).
The detailed annotation results
can be found in Section~\ref{sec:4.2.3}.
%are shown in Table \ref{table:rq2_dataset}.\lin{This table is too late, how to fix} 

% \jie{here, we may need to explicit state which categories are syntactic noise and which categories are semantic noise.}\lin{updated}

%  \begin{figure*}[]
% \centering
%  \includegraphics[width=\textwidth,height=21.5cm]{fig/taxonomy-tab.pdf}
%  %\vspace{-0.3cm}
%  \label{tab:tax-tmp}
%  \end{figure*}

\subsection{Comment-Related Noisy Data}
\textbf{Partial Sentence.}
{Since it is a common practice to place a method's summary at the first sentence of its comment~\cite{technetwork_articles_java}, most researchers use the first sentences of the code comments as the target summaries. While, we have observed that some inappropriate processing can lead to partial first sentences collected. For example, {\funcom} only collects the first line from the following java doc as the comment, \textit{i.e.}, ``Returns the high value'', where the next line that should be part of the first sentence is missing. This is primarily due to automatic splitting using new line characters such as ``$\backslash$n''.
%Taking the partial first sentence as the target summary will lose some information, making the ground-truth unable to summarize the overall functionality of the source code. Besides, the partial first sentence will make the ground-truth summary shorter and thus easier to generate. 
%\jie{in this current experimental results, we'd better delete this sentence. }
}

\begin{lstlisting}
/* Returns the high-value
 * for an item within a series. */
\end{lstlisting}
\vspace{-2mm}
\begin{small}
\texttt{Comment ({\funcom}): returns the high value}
\end{small}
\vspace{3pt}

\noindent \textbf{Verbose Sentence.} 
%When collecting the first sentence as the short summary, researchers use `period' as the unique delimiter. However, developers may write their summaries across multiple lines, 
When collecting the first sentence as the target comment, some inappropriate processing  will lead to verbose first sentences collected. For example, {\pcsd} excessively includes the argument description ``arguments course data'' into the functionality summary.
\begin{lstlisting}
"""
Generate a CSV file containing a summary of the xBlock usage
Arguments:course_data
"""
\end{lstlisting}
\vspace{-2mm}
\begin{small}
\texttt{Comment ({\pcsd}): generate a csv file containing a summary of the xblock usage \textcolor{red}{arguments course data}}
\end{small}

\vspace{3pt}
\noindent \textbf{Content Tampering.} Developers may use HTML tags for documentation auto-generation or URLs for external references in comments. We observe that some inappropriate processing will keep the tags or URL contents together with the comments, thus contaminating the benchmark data with meaningless text. 
For example, {\csn} reserves the HTML tag ``p'' at the beginning and end of the comment.

\begin{lstlisting}
/* <p> Builds the JASPIC application context.</p> */
 \end{lstlisting}
 \vspace{-2mm}
 \begin{small}
\texttt{Comment ({\csn}): \textcolor{red}{p} builds the jaspic application context \textcolor{red}{p}}
 \end{small}
\vspace{3pt}
 
\noindent \textbf{Over-Splitting of Variable Identifiers.} Code comments are likely to contain variable identifiers or API terms when describing code functionalities. Splitting code by camelCase or snake\_case is a common operation for code understanding \cite{leclair2019neural,DBLP:conf/iwpc/ShahbaziSF21,DBLP:conf/acl/HasanMIMHHAIS21}. However, we observe that some studies perform this operation on {every matched token in the} comments including the predefined variable identifiers or API terms. 
For example, {\funcom} splits {a variable named} ``jTextField'' into ``j text field'' when collecting comments.
We consider such an operation can change the original meaning of code comments. 

\begin{lstlisting} 
/* This method initializes jTextField. */ 
\end{lstlisting}
\vspace{-2mm}
\begin{small}
\texttt{Comment ({\funcom}): this method initializes \textcolor{red}{j text field}}
\end{small}
\vspace{3pt}

\noindent\textbf{Non-Literal.} Developers from different countries may write comments in their first languages, mixing with the English language in the comments sometimes. We observe that existing benchmark datasets occasionally discard the Non-English text but remain the English text as code comments.
For example, {\csn} only extracts the English words, \textit{i.e.,} ``jsonarray bean list arraylist'' {from the following mixed comment that contains both Chinese and English words} as the summarization for the corresponding source code. 
Since the remaining comment data are typically incomplete and meaningless, we consider them as noises.

\vspace{3mm}

\begin{small}
% \begin{CJK*}{UTF8}{gbsn}
% \color[HTML]{009A17}{\makecell[l]{\ \texttt{/*}\\ \ \texttt{*}\texttt{\ 必须存在这个key的值, 不然抛IllegalStateException异常}\\ \ \texttt{*/}\\}}
% \end{CJK*}
\begin{CJK*}{UTF8}{gbsn}
\color[HTML]{009A17}{\makecell[l]{\  \colorbox{backcolour}{\texttt{/*\ 将JSONArray转换为Bean的List，默认为ArrayList\  */    }}\\}}
\end{CJK*}
\vspace{-2mm}
\texttt{Comment ({\csn}): jsonarray bean list arraylist}
\end{small}

\vspace{3mm}
\vspace{3pt}

\noindent \textbf{Interrogation.}
Based on our observation, some of the comments in the benchmark dataset are interrogations.
For example, in {\csn}, the comment for the \textit{isDue()} method is ``do we need to show the upgrade wizard prompt''.
%, e.g., 'Is it a good idea using USERID from ASPNET membership table as a foreign key?'. 
Such interrogations are mainly used for communication, rather than summarizing functionalities.

\begin{lstlisting}
/* Do we need to show the upgrade wizard prompt? */
public boolean isDue() {
  if (isUpToDate)
     return false; ...
 \end{lstlisting}
    \vspace{-2mm}
\begin{small}
\texttt{Comment ({\csn}): do we need to show the upgrade wizard prompt ?}
\end{small}
\vspace{2pt}

%Therefore, we remove such noises.

%Developers seem to use comments to communicate with their collaborators during the code review process. There may be some sparse information about the code functionality, but the quality is hard to control. 

\noindent \textbf{Under-Development Comments.} Based on our observation, some of the comments are related to ongoing and future development, including temporary tips, notes, \textit{etc}. For example, {\tlc} has a comment ``description of the method'' for the \textit{openFile} method, which is of little worth for understanding code. 
Since the under-development comments are typically inappropriate for the scenario of automated code summarization, we consider them as noises.
 %envelopes temporary tips, notes, or suggestions that developers use during development

\begin{lstlisting}
/* Description of the Method */
protected void openFile(File f) {
  if (f == null) { ...
\end{lstlisting}
\vspace{-2mm}
\begin{small}
\texttt{Comment ({\tlc}): description of the method}
\end{small}
% \vspace{2.5mm}

%\textit{For the code data in the benchmark datasets, we identify the following five categories of syntactical noises:}

\subsection{Code-Related Noisy Data}

\textbf{Empty Function.} Developers often take on technical debt to speed up software development~\cite{wehaibi2016examining}. It has been widely observed that empty function is a common type of technical debt. However, the code-comment pairs extracted from these empty functions can introduce non-trivial noises, this is because an unimplemented empty function and its comment do not match either syntactically or semantically. 
For example, {\funcom} includes an empty method \textit{end()} with a 10-word comment.

% \begin{lstlisting}
% /* Log an error of the specified category. */
% public void logError(CacheErrorCategory category){}
%  \end{lstlisting}
%     \vspace{-2mm}
% \begin{small}
% \texttt{Code in CSN: public void log error cache error category category}
% \end{small}
% \vspace{2.5mm}

\begin{lstlisting}
/*Specifies the behaviour of the automaton in its end state*/
protected void end(){}
 \end{lstlisting}
    \vspace{-2mm}
\begin{small}
\texttt{Code ({\funcom}): protected void end}
\end{small}
\vspace{2.5mm}

%We remove the code-summary pairs extracted from empty function for data purification.

\textbf{Commented-Out Method.} Developers often comment out a whole method for deprecating a specific functionality~\cite{falessi2015five}. We observe that, in the studied benchmark datasets, some commented-out methods are collected as the comments for the sequential methods. 
For example, the commented-out method \textit{transformTypeID} and its comments are still included in {\funcom}.
%We consider these code-summary pairs extracted from commented-out methods as noises and remove them for cleaning data. 

\begin{lstlisting}
/* for now try mappig full type URI  */
// public String transformTypeID(URI typeuri){
// return typeuri.toString();}
 \end{lstlisting}
   \vspace{-2mm}
\begin{small}
\texttt{Code ({\funcom}): public string transform type id ...}
\end{small}
\vspace{2.5mm}

\textbf{Block-Comment Code.} We have observed that some code in the benchmark datasets contains block comments inside their bodies. The blocked comments could be natural-language comments or commented-out code. 
For example, the block comment ``TODO: Why is he using Math.round'' is considered as a piece of code for the \textit{getFixQuality} method in {\funcom}.
If keeping these blocked comments in the source code, the original logics of the code are likely to be distorted when tokenizing it for code summarization models.

\begin{lstlisting}
/* Get GPS Quality Data  */
public int getFixQuality(){
  checkRefresh();
  // TODO: Why is he using Math.round?
  Return Math.round(quality);}
 \end{lstlisting}
  \vspace{-2mm}
\begin{small}
\texttt{Code ({\funcom}): public int get fix quality check refresh \textcolor{red}{todo why is he using math round} return math round quality}
\end{small}
\vspace{2.5mm}

\textbf{Auto Code.} Developers often use modern IDEs like Eclipse or Visual Studio to generate auxiliary functions such as \textit{getter}, \textit{setter}, \textit{toString}, or \textit{tester} for some predefined variables. 
The comments for these {auto generated methods} are often  similar to or the same as the method names, which makes the code-comment pairs less informative. 
{For example, in {\funcom}, the comment for the auto-generated test method (\textit{i.e.}, \textit{testConstructor}) is ``Test the constructor'', which is almost the same as the method name after splitting.} 
%For example, in {\tlc}, the auto-generated comment ``Auto generated test method'' exists multiple places where the corresponding code are different.

\begin{lstlisting}
/* Test the constructor */
public void testConstructor() {
  System TestResult str;
  System TestID testID1; ...
\end{lstlisting}
\vspace{-2mm}
\begin{small}
\texttt{Comment ({\funcom}): test the constructor}\\
\texttt{Code ({\funcom}): public void test constructor ...}
\end{small}
\vspace{2.5mm}

\textbf{Duplicated Code.} 
Developers often reuse code by copying, pasting and modifying to speed up software development~\cite{DBLP:conf/oopsla/Allamanis19, DBLP:conf/icse/SajnaniSSRL16}. 
These code snippets often have similar or the same comments. Sharing identical code and summarization pairs in the training and test sets is inappropriate and would make the model learn these cases easily.
%\rev{Performance of models trained on these data can decline significantly on a different dataset that does not contain the duplicated code.} \lin{last sentence reads wired.}
%regarding BLEU or ROUGE.
%``real'' examples while falsely inflating performance metrics such as BLEU or Rouge.

%Lige: 没太搞懂这一类的意思。是说，因为 Duplicated codes 的summarization总是相同的，所以机器学习的方法容易背出来，所以会影响BLEU吗？

\section{The Code-Comment Cleaning Tool}
\label{sec:cleaningtool}
 
{To support automatic detection {of noises in} the proposed taxonomy, we develop a code-comment cleaning tool, named {\tool}, based on a set of heuristic rules. 
This section introduces the design of the rule-based cleaning tool, and presents the analysis results of its effectiveness.}
%As shown in Figure \ref{fig:overview}, the tool consists of two subsequent cleaners, a rule-based syntactic cleaner and a model-based semantic cleaner. First, the rule-based syntactic cleaner is used for removing syntactic noises, and then the model-based semantic cleaner is used for classifying non-summary and summary comments from the rest (Note that, we refer to semantic noises as non-summary comments and refer to normal comments as summary comments in this study).\lin{updated @Ge} 
%The origin code-summary pairs in benchmark datasets are firstly cleaned by the rule-based cleaner, where a set of heuristic rules are applied to detect the existence of invalid syntax.  Next, we further filter semantic noise based on our model-based cleaner, which is trained on an external dataset.

\subsection{The Heuristic Rules}
\begin{table}[b]
%\vspace{-0.3cm}
\caption{Syntax features and our actions in heuristic rules}
%\vspace{-0.2cm}
\label{tab:taxonomy}
\centering
\resizebox{\columnwidth}{!}{
% \scalebox{0.7}{
% \setlength\tabcolsep{2pt} 
\begin{tabular}{|cc|c|c|}
\hline
\multicolumn{2}{|c|}{\textbf{Category}}                                             & \textbf{Syntax Feature}                                                                & \textbf{Action} \\ \hline \hline
\multicolumn{1}{|c|}{\multirow{14}{*}{\rotatebox{90}{\textit{Comment}}}} & \makecell{Partial  Sentence}     & \makecell{Shorter than the corrected\\ first sentence}          & \makecell{UPDATE: Replace with\\ the corrected\\ first sentence } \\ \cline{2-4} 
 \multicolumn{1}{|c|}{}                         & \makecell{Verbose Sentence}     & \makecell{Longer than the corrected\\ first sentence}                                                       & \makecell{UPDATE: Replace with\\ the corrected\\ first sentence } \\ \cline{2-4} 
 \multicolumn{1}{|c|}{}                         & \makecell{Content Tampering}    & \makecell{HTML tags, Doc tags, \\and URL format}  & \makecell{UPDATE: Clean the tags\\ from comment data}\\ \cline{2-4} 
 \multicolumn{1}{|c|}{}                         & Over-Splitting       &\makecell{Split comments on camel\\ case and underscore}                                   & \makecell{UPDATE: Replace the \\over-splitting variables\\ with the original ones } \\ \cline{2-4} 
 \multicolumn{1}{|c|}{}                         & Non-Literal          & non-ASCII                                                                                 & \makecell{REMOVE} \\ \cline{2-4} 

 \multicolumn{1}{|c|}{}                         & Interrogation        & \makecell{``?'', ``what'', ``how'', \textit{etc}. }                      & \makecell{REMOVE} \\ \cline{2-4} 
 \multicolumn{1}{|c|}{}                         & Under-Development          & \makecell{``todo'', ``deprecate'',\\ ``copyright'',``FIXME:'', \textit{etc}.}       & \makecell{REMOVE} \\ \hline \hline
 \multicolumn{1}{|c|}{\multirow{9}{*}{\rotatebox{90}{\textit{Code}}}}    & \makecell{Empty  Function}       & The method body is empty                                                                & \makecell{REMOVE} \\ \cline{2-4} 
 \multicolumn{1}{|c|}{}                         & \makecell{Commented-Out Method} & \makecell{The whole method \\is commented out.}             & \makecell{REMOVE} \\ \cline{2-4} 
 \multicolumn{1}{|c|}{}                         & \makecell{Block-Comment Code}   & \makecell{The method contains the\\ block comment.}                                                                                                         & \makecell{UPDATE: Clean the \\blocked comments\\ from the code body} \\ \cline{2-4} 
 \multicolumn{1}{|c|}{}                         & Auto code            & \makecell{setter, getter, tester, \textit{etc}.}                                                                           & \makecell{REMOVE} \\ \cline{2-4} 
 \multicolumn{1}{|c|}{}                         & \makecell{Duplicated \\Code}      & Exact Match                                  & \makecell{REMOVE} \\ \hline
 
\end{tabular}
}
\end{table}
%According to common practice guidance, e.g., Javadoc guidance \cite{}, 
%http://www.oracle.com/technetwork/articles/java/index-137868.html
%the first sentences of the code comments typically describe the code functionalities. 
%Therefore, nearly all the existing researches extract the first sentences of the code comments as the target summaries. 

%Since raw data can hardly be directly used for code summarization tasks, nearly all the existing benchmark datasets are constructed under multiple preprocessing operations, e.g., using `newline' as the delimiter to collect the first sentence in the comments \cite{leclair2019neural}.
%\jie{not only the preprocessing, but also the data itself, cause the noise in the data. i think this current statement might be misleading.}
%However, we have a concern that about how well are these preprocessing operations performed in constructing benchmark datasets. 
%To address the above concern, we develop a syntactic cleaner with a set of heuristic rules that can precisely identify syntactic noises.

%\subsubsection{Taxonomy of syntactic noises.} 

%We introduce the taxonomy of syntactic noises and their examples in Table XXX.
%To automatically detect each category of noisy data in the taxonomy, we develop a cleaner with a set of heuristic rules.

\textbf{Construction criteria.}
The heuristic rules conform to the following criteria: (1) Each rule should define a unique and specific category without overlap; (2) Rules should limit the exclusion of valid data within an acceptable range, {\textit{i.e.}, all the F1 scores should be larger than 90\%}; %\song{what does this mean?};
and (3) Any rule is not a subrule of the others. 

%\jie{the reviewers might argue what is the `acceptable range?'}\lin{I follow previous study when state this, our inspection shows all above 90\% F1 score, any idea how can we update here @junjie}\jie{i am not sure}

% Algorithm 1 illustrates the details of the syntactic cleaner detecting and correcting Partial sentence and Verbose sentences. Initially, we set the first sentence of raw comment $FS$ with an empty string and define three regex.
% Then, we split the input raw comment $RC$ into a comment list $L$ using newline character (5-th lines). We loop through each comment line $l$ in the comment List $L$ until the first sentence is extracted (6-th to 17-th lines).  Specifically, at each loop time, we first remove the comment tags such as `*' at start of the comment line $l$ by using regex1 (12-th lines). We then use the regex2 to determine whether the current comment line $l$ contains a complete sentence. If not, we splice the current comment line $l$ to the end of the existing first sentence $FS$ (13-th to 17-th lines). If the existing $FS$ is not empty, we use regex3 to determine whether the current comment line l starts with an uppercase letter or special tags, and if so, we take the existing $FS$ string as the first sentence and break the loop (8-th to 9-th lines). Finally, we determine the noise category of the benchmark comment $BC$ by comparing its length with the extracted first sentence $FS$ (18-th to 22-th lines).
%\input{table/alg}

\begin{figure}[t]
\centering
\vspace{-0.1cm}
\includegraphics[width=0.75\columnwidth]{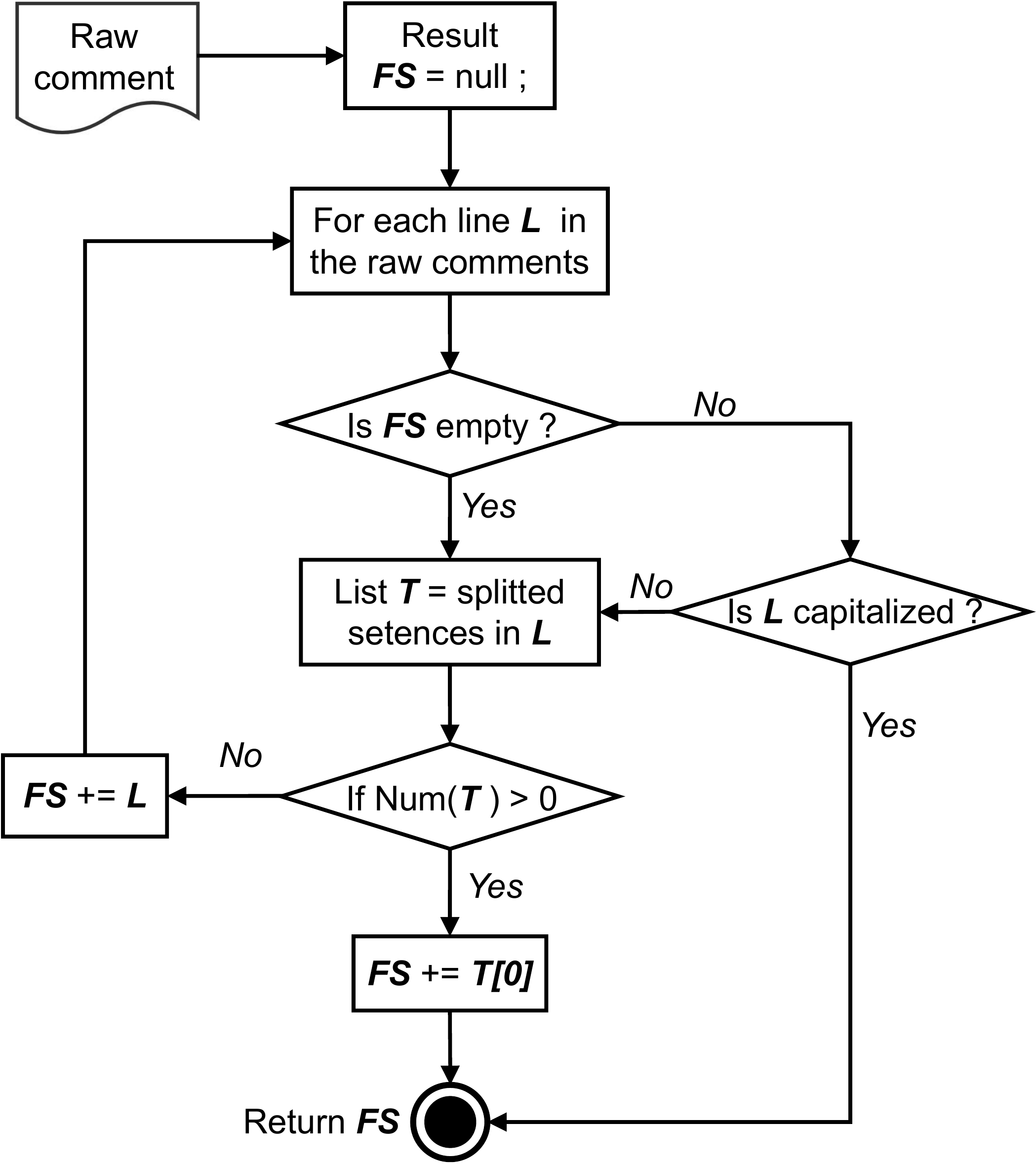}
\vspace{-0.3cm}
\caption{If-else rules of collecting the first sentence in the raw comments for partial and verbose sentence noises.}
\vspace{-0.4cm}
\label{fig:3-flow}
\end{figure}

\textbf{Construction process.}
For each category of noisy data, we develop a set of if-else rules to detect them by the following steps.
(1) Based on the manually annotated noises produced in Section~\ref{sec:tox}, we carefully identify syntax features for each category from 80\% of the manual data; 
(2) We design a set of if-else rules to {detect} 
the noisy data from the raw data (note that, the raw data refers to the source data that has not been processed for use, and the origin data refers to the processed raw data in the benchmark datasets);
(3) To avoid overfitting, we test the correctness of the rules on the rest 20\% of the manually annotated noises. 
We iteratively adjust the rules until the performance is acceptable, \textit{i.e.}, over 90\% F1 score.

\textbf{Example Rules.}
Figure \ref{fig:3-flow} 
illustrates the if-else rules of collecting the first sentence in the raw comments for partial and verbose sentence noises. The key idea is to sequentially determine whether each line of comment in a raw comment contains a complete sentence. If so, return the first complete sentence; if not, save the content of the line and continue to determine the next line. 
By comparing the first sentence we extracted from raw data with the processed sentence provided in the benchmark datasets, we can determine the verbose or partial sentence category.
%Table \ref{tab:taxonomy} shows the syntax features and our actions for each category.

Table \ref{tab:taxonomy} demonstrates the syntax feature of heuristic rules and our actions {to resolve noises detected}.
Details of the implementation of each category can be found on our website \cite{website}.

\subsection{Effectiveness Evaluation}
%In this section, we evaluate the effectiveness of our code-comment cleaning tool.

\subsubsection{Data Preparation.}
As introduced in Section~\ref{sec:3.1}, we labeled 12 categories of noisy data from 1,600 code-comment pairs sampled from the four benchmark datasets. 
These manually annotated data are used to evaluate the performance of CAT. 
We build the heuristic rules based on observing 80\% of the annotated noisy data, and evaluate on the rest 20\%. 
The ``Dataset'' column in Table \ref{table:rq2_dataset} shows the detail.

% \subsubsection{Experimental setup}
% We implement the model-based semantic cleaner using scikit-learn \cite{scikitlearn} which is an open-source data analysis and machine learning library. We train the GBDT model with log-likelihood loss function. 
% {For the} hyper-parameters,
% we use greedy search as the parameter tuning  method to
% obtain the best performance.
% {After the parameter tuning, we set the \textit{learning\_rate}=0.01, \textit{n\_estimators}=100, \textit{subsample}=1, and \textit{min\_samples\_leaf}=2.}

\subsubsection{Evaluation Metrics.}

We use three commonly-used metrics to evaluate the performance of CAT, \textit{i.e.}, \textit{Precision}, \textit{Recall}, and \textit{F1}. (1) \textit{Precision} refers to the ratio of correct predictions to the total number of predictions; (2) \textit{Recall} refers to the ratio of correct predictions to the total number of samples in the golden test set; and (3) \textit{F1} is the harmonic mean of precision and recall. 
%When comparing the performances, we care more about F1 since it is balanced for evaluation.

\begin{table}[b]
\vspace{-0.3cm}
\caption{Effectiveness of noise detection rules}
\vspace{-0.2cm}
\label{table:rq2_dataset}
\resizebox{\columnwidth}{!}{
\begin{tabular}{|cc|ccc|ccc|}
\hline
\multicolumn{2}{|c|}{\multirow{4}{*}{\textbf{Category}}}                       & \multicolumn{3}{c|}{\textbf{Dataset}}                                                                                                                                                                                                                           & \multicolumn{3}{c|}{\textbf{Performance (\%)}}                                       \\ \cline{3-8} 
\multicolumn{2}{|c|}{}                                                         & \multicolumn{1}{c|}{\textbf{\begin{tabular}[c]{@{}c@{}}\makecell{\#Anno-\\tations\\ (100\%)}\end{tabular}}} & \multicolumn{1}{c|}{\textbf{\begin{tabular}[c]{@{}c@{}}\makecell{Rule-\\Build\\ (80\%)}\end{tabular}}} & \textbf{\begin{tabular}[c]{@{}c@{}}\makecell{Rule-\\Test\\ (20\%)}\end{tabular}} & \multicolumn{1}{c|}{\textbf{P}} & \multicolumn{1}{c|}{\textbf{R}} & \textbf{F1} \\ \hline \hline
\multicolumn{1}{|c|}{\multirow{8}{*}{\rotatebox{90}{\textit{Comment}}}} & \makecell{Partial Sentence}     & \multicolumn{1}{c|}{176}                                                                      & \multicolumn{1}{c|}{135}                                                                  & 41                                                                  & \multicolumn{1}{c|}{97.5}      & \multicolumn{1}{c|}{95.1}      & 96.3       \\ \cline{2-8} 
\multicolumn{1}{|c|}{}                                  & \makecell{Verbose Sentence}     & \multicolumn{1}{c|}{129}                                                                      & \multicolumn{1}{c|}{111}                                                                  & 18                                                                  & \multicolumn{1}{c|}{94.7}      & \multicolumn{1}{c|}{100.0}        & 97.3       \\ \cline{2-8} 
\multicolumn{1}{|c|}{}                                  & \makecell{Content Tampering}    & \multicolumn{1}{c|}{147}                                                                      & \multicolumn{1}{c|}{120}                                                                  & 27                                                                  & \multicolumn{1}{c|}{92.9}      & \multicolumn{1}{c|}{96.3}      & 94.6       \\ \cline{2-8} 
\multicolumn{1}{|c|}{}                                  & Over-Splitting       & \multicolumn{1}{c|}{84}                                                                       & \multicolumn{1}{c|}{63}                                                                   & 21                                                                  & \multicolumn{1}{c|}{90.9}      & \multicolumn{1}{c|}{95.2}      & 93.0       \\ \cline{2-8} 
\multicolumn{1}{|c|}{}                                  & Non-Literal          & \multicolumn{1}{c|}{38}                                                                       & \multicolumn{1}{c|}{30}                                                                   & 8                                                                   & \multicolumn{1}{c|}{100.0}        & \multicolumn{1}{c|}{100.0}        & 100.0         \\ \cline{2-8} 
\multicolumn{1}{|c|}{}                                  & Interrogation        & \multicolumn{1}{c|}{16}                                                                       & \multicolumn{1}{c|}{7}                                                                    & 9                                                                   & \multicolumn{1}{c|}{100.0}        & \multicolumn{1}{c|}{88.9}      & 94.1       \\ \cline{2-8} 
\multicolumn{1}{|c|}{}                                  & \makecell{Under-Development}    & \multicolumn{1}{c|}{57}                                                                       & \multicolumn{1}{c|}{92}                                                                   & 57                                                                  & \multicolumn{1}{c|}{91.5}      & \multicolumn{1}{c|}{94.7}      & 93.1       \\ \cline{2-8} 
\multicolumn{1}{|c|}{}                                  & \textit{Total}       & \multicolumn{1}{c|}{647}                                                                      & \multicolumn{1}{c|}{558}                                                                  & 181                                                                 & \multicolumn{1}{c|}{95.4}      & \multicolumn{1}{c|}{95.8}      & 95.5       \\ \hline \hline
\multicolumn{1}{|c|}{\multirow{6}{*}{\rotatebox{90}{\textit{Code}}}}    & Empty Function       & \multicolumn{1}{c|}{21}                                                                       & \multicolumn{1}{c|}{14}                                                                   & 7                                                                   & \multicolumn{1}{c|}{100.0}        & \multicolumn{1}{c|}{100.0}        & 100.0         \\ \cline{2-8} 
\multicolumn{1}{|c|}{}                                  & \makecell{Commented-Out Method} & \multicolumn{1}{c|}{4}                                                                        & \multicolumn{1}{c|}{2}                                                                    & 2                                                                   & \multicolumn{1}{c|}{100.0}        & \multicolumn{1}{c|}{100.0}        & 100.0         \\ \cline{2-8} 
\multicolumn{1}{|c|}{}                                  & \makecell{Block-Comment Code}   & \multicolumn{1}{c|}{44}                                                                       & \multicolumn{1}{c|}{31}                                                                   & 13                                                                  & \multicolumn{1}{c|}{100.0}        & \multicolumn{1}{c|}{92.3}      & 96.0          \\ \cline{2-8} 
\multicolumn{1}{|c|}{}                                  & Auto Code            & \multicolumn{1}{c|}{179}                                                                      & \multicolumn{1}{c|}{133}                                                                  & 46                                                                  & \multicolumn{1}{c|}{97.7}      & \multicolumn{1}{c|}{93.5}      & 95.6       \\ \cline{2-8} 
\multicolumn{1}{|c|}{}                                  & Duplicated Code      & \multicolumn{1}{c|}{22}                                                                       & \multicolumn{1}{c|}{16}                                                                   & 6                                                                   & \multicolumn{1}{c|}{100.0}        & \multicolumn{1}{c|}{100.0}        & 100.0         \\ \cline{2-8} 
\multicolumn{1}{|c|}{}                                  & \textit{Total}       & \multicolumn{1}{c|}{270}                                                                      & \multicolumn{1}{c|}{196}                                                                  & 74                                                                  & \multicolumn{1}{c|}{99.6}      & \multicolumn{1}{c|}{97.2}      & 98.3       \\ \hline
\end{tabular}
}
\vspace{-0.2cm}
\end{table}
\subsubsection{Results}
\label{sec:4.2.3}

Table \ref{table:rq2_dataset} demonstrates the performance of {\tool}. We can see that, it can accurately detect noises on the four benchmark datasets. The F1 scores of detecting comment-related noises are ranging from 93.0\% to 100.0\%,  and 95.5\% on average.
The average F1 scores of detecting code-related noises are ranging from 95.6\% to 100.0\%,  and 98.3\% on average.
%The accuracy of syntactic noise correction are ranging from 92.06\% to 95.18\%, which is also highly satisfactory. 
The results show that, {\tool} can achieve highly satisfactory performance on filtering noisy data from code-comment datasets.
In summary, our code-comment cleaning tool can 
{accurately} filter noisy data, with all the F1 scores of over 90.0\%, {which can help build a high-quality dataset for the follow-up code summarization tasks.}
%I think we should add a sentence like that.  }

% \noindent
% \begin{boxedminipage}[t]{\columnwidth}
% \jie{i think it isn't necessary to treat this as a separate finding. this is not a finding. }
% \textbf{Finding 1}:   
% \end{boxedminipage}

% \vspace{-0.1cm}
\section{Quality Assessment of Benchmarks}
\label{sec:qassessment}
% \begin{figure*}[htpb]
% \centering
% \includegraphics[width=1.7\columnwidth]{fig/pie.pdf}
% %\vspace{-0.3cm}
% \caption{Percentage of syntactic (SYN) and semantic (SEM) noise in terms of removal and update.\lin{add overall}}
% \label{fig:rq1-pie}
% \end{figure*}

% \usepackage{multirow}
\vspace{-0.1cm}
\begin{table}[b]
\caption{Distribution of noisy data in benchmark datasets.}
\vspace{-0.2cm}
\label{tab:heat}
\centering
\resizebox{\columnwidth}{!}{
\begin{tabular}{|cc|c|c|c|c|}
\hline
\rowcolor[HTML]{FFFFFF} 
\multicolumn{2}{|c|}{\textbf{Category of Noisy Data }}   & \textbf{\funcom\ (\%)}                         & \textbf{\ \ \tlc\ (\%)\ \ \ }                            & \textbf{\ \ \codenn\ (\%)\ \ \ }                            & \textbf{\ \ \pcsd\ (\%)\ \ }                         \\ \hline
\multicolumn{2}{|c|}{\textit{Total}}  &
\hspace{-1.3cm}
\begin{minipage}{0cm}
    \includegraphics[width=1.1cm,height=1.1cm]{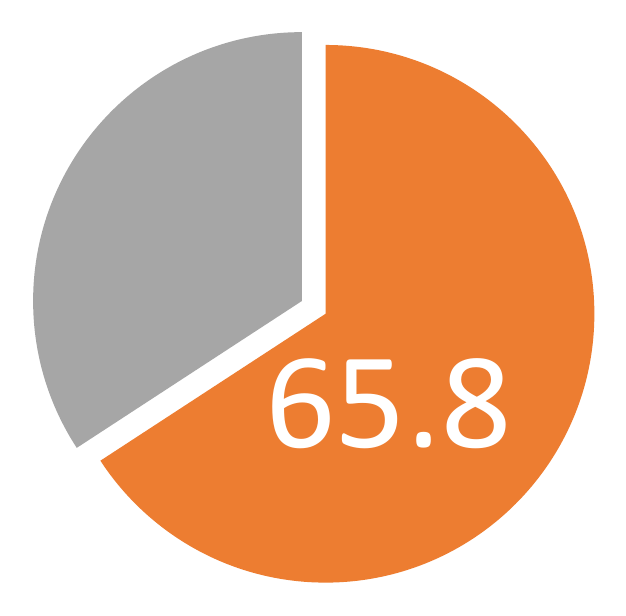}
\end{minipage} &
\hspace{-1.3cm}
\begin{minipage}{0cm}
    \includegraphics[width=1.1cm,height=1.1cm]{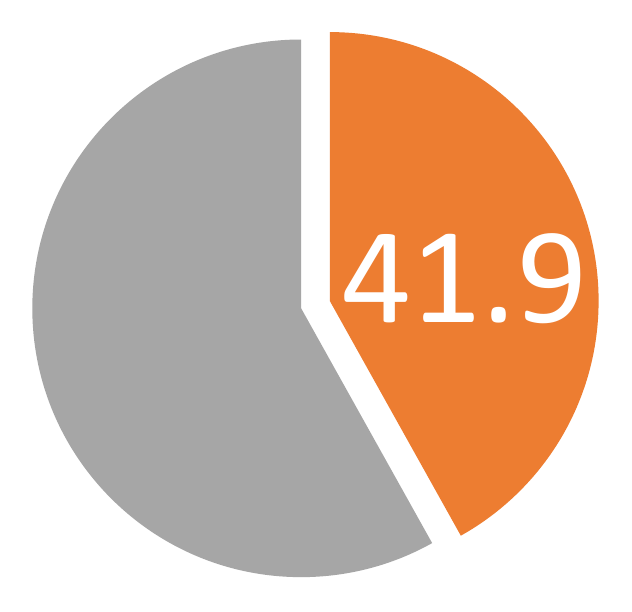}
\end{minipage}&\hspace{-1.3cm}
\begin{minipage}{0cm}
    \includegraphics[width=1.1cm,height=1.1cm]{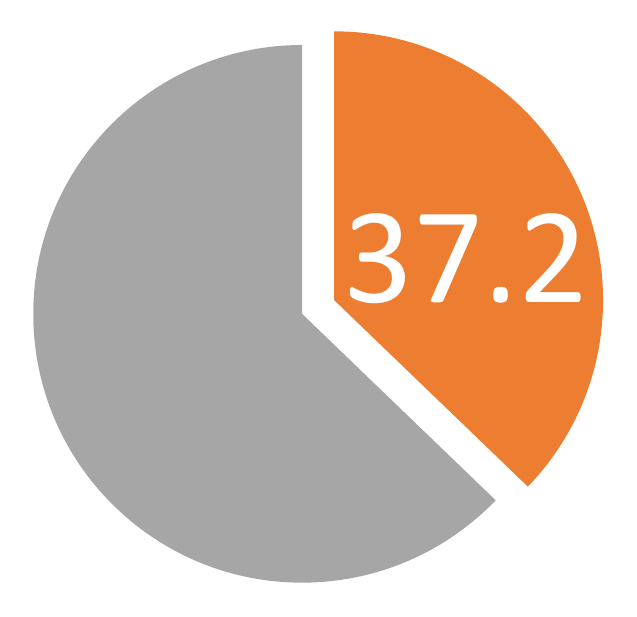}
\end{minipage}&\hspace{-1.3cm}
\begin{minipage}{0cm}
    \includegraphics[width=1.1cm,height=1.1cm]{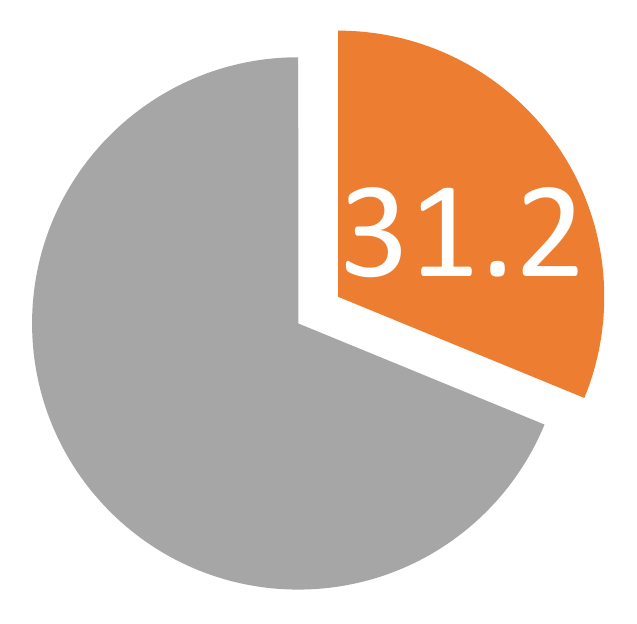}
\end{minipage}
\\
\hline \hline
\multicolumn{1}{|c|}{\multirow{8}{*}{\rotatebox{90}{\textit{Comment}}}}                         & \makecell{Partial Sentence}     & \cellcolor[HTML]{D78070}17.1 & \cellcolor[HTML]{F1D99E}0.0  & \cellcolor[HTML]{DC8E78}7.8  & \cellcolor[HTML]{D78271}15.9 \\ \cline{2-6} 
\multicolumn{1}{|c|}{}                          & \makecell{Verbose Sentence}     & \cellcolor[HTML]{F1D99E}0.0  & \cellcolor[HTML]{D3776B}22.8 & \cellcolor[HTML]{F1D99E}0.0  & \cellcolor[HTML]{DC8E78}7.8  \\ \cline{2-6} 
\multicolumn{1}{|c|}{}                          & \makecell{Content Tampering}    & \cellcolor[HTML]{DB8B77}9.7  & \cellcolor[HTML]{DE967C}3.2  & \cellcolor[HTML]{D27469}24.4 & \cellcolor[HTML]{E7B38B}0.5  \\ \cline{2-6} 
\multicolumn{1}{|c|}{}                          & \makecell{Over-Splitting}       & \cellcolor[HTML]{D3756A}24.1 & \cellcolor[HTML]{F1D99E}0.0  & \cellcolor[HTML]{F1D99E}0.0  & \cellcolor[HTML]{F1D99E}0.0  \\ \cline{2-6} 
\multicolumn{1}{|c|}{}                          & \makecell{Non-Literal}          & \cellcolor[HTML]{E6B28B}0.5  & \cellcolor[HTML]{F1D99E}0.0  & \cellcolor[HTML]{DC8E78}7.8  & \cellcolor[HTML]{EDCB97}0.2  \\ \cline{2-6} 
\multicolumn{1}{|c|}{}                          & \makecell{Interrogation}        & \cellcolor[HTML]{E2A182}0.7  & \cellcolor[HTML]{DF997E}0.9  & \cellcolor[HTML]{E2A182}0.7  & \cellcolor[HTML]{EABE91}0.3  \\ \cline{2-6} 
\multicolumn{1}{|c|}{}                          & \makecell{Under-Development}    & \cellcolor[HTML]{DE957C}3.7  & \cellcolor[HTML]{DF997E}1.2  & \cellcolor[HTML]{DF997E}1.2  & \cellcolor[HTML]{DF977D}2.3  \\ \cline{2-6} 
\multicolumn{1}{|c|}{}& \textit{Total}       & 40.9                         & 25.4                        & 36.1                         & 26.5                         \\ \hline \hline 
\multicolumn{1}{|c|}{\multirow{6}{*}{\rotatebox{90}{\textit{Code}}}}                          & \makecell{Empty Function}       & \cellcolor[HTML]{DF987E}1.6  & \cellcolor[HTML]{DF997E}1.1  & \cellcolor[HTML]{F1D89E}0.0  & \cellcolor[HTML]{F1D99E}0.0  \\ \cline{2-6} 
\multicolumn{1}{|c|}{}                          & \makecell{Commented-Out Method} & \cellcolor[HTML]{ECC795}0.2  & \cellcolor[HTML]{F1D99E}0.0  & \cellcolor[HTML]{F1D99E}0.0  & \cellcolor[HTML]{F1D99E}0.0  \\ \cline{2-6} 
\multicolumn{1}{|c|}{}                          & \makecell{Block-Comment Code}   & \cellcolor[HTML]{DA8975}11.1 & \cellcolor[HTML]{F1D99E}0.0  & \cellcolor[HTML]{F1D99E}0.0  & \cellcolor[HTML]{F1D99E}0.0  \\ \cline{2-6} 
\multicolumn{1}{|c|}{}                          & \makecell{Auto Code}            & \cellcolor[HTML]{CF6B64}29.8 & \cellcolor[HTML]{DD947B}4.6  & \cellcolor[HTML]{DF987E}1.6  & \cellcolor[HTML]{DE947B}4.3  \\ \cline{2-6} 
\multicolumn{1}{|c|}{}                          & \makecell{Duplicated Code}      & \cellcolor[HTML]{E5AE89}0.6  & \cellcolor[HTML]{D67E6F}18.4 & \cellcolor[HTML]{F1D99E}0.0  & \cellcolor[HTML]{DF987E}1.5  \\ \cline{2-6} 
\multicolumn{1}{|c|}{}    & \textit{Total}       & 40.7                         & 22.6                         & 1.6                          & 5.8                          \\ \hline \hline
\multicolumn{2}{|c|}{\textit{Removed noisy data}}                               & 38.7                         & 21.1                         & 29.2                        & 9.3                         \\ \hline
\multicolumn{2}{|c|}{\textit{Updated noisy data}}                               & 27.1                         & 20.8                         & 8.0                          & 21.9                         \\ \hline
\end{tabular}
}
\vspace{-0.3cm}
\end{table}

In this step, the code-comment cleaning tool is applied to detect and correct noises through comment removal or update actions as listed in Table 2. Based on the noisy data output by the tool, we further analyze the quality of the four benchmark datasets.
Table~\ref{tab:heat} illustrates the distribution of each noise category on the four benchmark datasets. 
The number on each cell presents the percentage of the noises in the corresponding benchmark dataset, directly generated from the cleaning tool. Note that, since one code-comment pair may involve multiple noises, the tool repeatedly counts those that involve multiple noise categories when calculating frequency for each category, and counts once for the total frequency. Thus, the sum of individual category percentages is slightly higher than the percentage of total noises.

%#categories
\textbf{Overall,} 
{\funcom} has the highest proportion of noisy data (65.8\%), followed by {\tlc} (41.9\%), {\csn} (37.2\%), and {\pcsd} (31.2\%).
We can also observe that, the benchmark datasets often contain multiple categories of noises.
{\funcom} contains the most noise categories. Except for the verbose sentence noises, every other category is included. The other three benchmark datasets contain seven or eight categories. 

%The frequency of total noises is range from 31\% to 66\% among the four benchmark datasets, and the average frequency is 44\%. Among them, Funcom is the most noisy dataset with 62\% syntactic and 4\% semantic noisy code-comment pairs, while PCSD is the least noisy dataset with 29\% syntactic and 2\% semantic noisy code-comment pairs. 

%comment noise
\textbf{Noise distribution in comments}. It is observed that %we can see that, 
40.9\% comments in {\funcom} contain noises, followed by {\csn}, {\tlc}, and {\pcsd}.
Specifically, all the four benchmark datasets have content-tampering, interrogation, and under-development noisy comments.
24.4\% comments in {\csn} {are contaminated by the meaningless text such as \textit{HTML} tags, \textit{Javadoc} tags, or \textit{URLs}.} 
24.1\% comments in {\funcom} are over-spliting by camelCase. 
22.8\% comments in {\tlc} are verbose sentences.

%code noise
\textbf{Noise distribution in source code.} It is also observed that 40.7\% source code in {\funcom} contain noises, followed by {\tlc}, {\pcsd}, and {\csn}. 
Specifically, all the four benchmark datasets have auto-code noises.
In {\funcom}, 29.8\% code is auto-generated such as \textit{setter}, \textit{getter}, \textit{tester}, and \textit{toString} methods.
Indeed, previous research~\cite{DBLP:conf/acl/HasanMIMHHAIS21} used to complain about similar issues that {{\funcom} contains much auto-generated code.}
In {\tlc}, 18.4\% code is exactly duplicated while the other three benchmark datasets are nearly none. This phenomenon indicates that preprocessing operations applied on existing benchmark datasets are not coincident all the time in that, some benchmark datasets apply the dedup preprocessing operation while some do not.

%Second, 24.06\% of the summaries are split unnecessarily by camelCase. The major noises for {\tlc} are that,  22.85\% summaries are verbose sentences, and 18.42\% code are duplicated. The noises in {\codenn} are concentrated in content tampering. 24.37\% summaries {are contaminated by the meaningless text such as \textit{HTML} tags, \textit{Javadoc} tags or \textit{URL}.} 
  
%The primary noises with {\pcsd} are that, 15.91\% summaries are the partial first sentences. 

\textbf{Distribution of updates and removals.}
The bottom part of Table~\ref{tab:heat} shows the frequency of different types of noise that were removed or updated from the four benchmark datasets based on {the corrective} actions introduced in Table \ref{tab:taxonomy}. 
{\funcom} and {\tlc} have high proportions of both removals and updates, \textit{i.e.}, 38.7\% and 27.1\% noisy data in {\funcom} are removed and %and 27.1\% data in {\funcom}
updated respectively. 
The major correction for {\csn} is removals. While the major correction for {\pcsd} is updating noises. This might be caused by the different noise distributions in these two benchmark datasets.

%In the {\tlc} and {\pcsd} datasets, the updated noises are more frequent than removal noises. The other two datasets ({\funcom} and {\codenn}) have a lower frequency of updated noises, especially the {\codenn} dataset has no updated noise. The reason is that the {\codenn} has only Interrogation and Duplicated Code noise, both of which are removal noise.

%  \begin{figure}[ht]
% \centering
%  \includegraphics[width=\columnwidth]{fig/heatmap.pdf}
%  %\vspace{-0.3cm}
%  \caption{Heatmap for Noise distribution (\%).}
%  \label{fig:rq1-heat}
%  \end{figure}

% \noindent
% \begin{boxedminipage}[t]{\columnwidth}
% \textbf{Finding 1}: Syntactic and semantic noises extensively exist in the four widely-used benchmark datasets, especially the syntactic noises ranging from 28.90\% to 62.08\%. 29.78\% of the code in {\funcom} are auto-generated; 22.84\% summaries in {\tlc} are verbose first sentences; 24.37\% summaries in {\codenn} are contaminated by the meaningless text; and 15.91\% summaries in {\pcsd} are the partial first sentences.
% \end{boxedminipage}

\begin{center}
\begin{tcolorbox}[colback=gray!10,%gray background
                  colframe=black,% black frame colour
                  width=8.3cm,% Use 8cm total width,
                  arc=1mm, auto outer arc,
                  boxrule=0.5pt,
                  left=3pt,
                  right=3pt,
                  top=3pt,
                  bottom=3pt
                 ]
\textbf{Finding 1}: Noisy data extensively exist in the four widely-used benchmark datasets, ranging from 31.2\% to 65.8\%. 29.8\% of the code in {\funcom} is auto-generated; 22.8\% comments in {\tlc} are verbose first sentences; 24.4\% comments in {\codenn} are contaminated by the meaningless text; and 15.9\% comments in {\pcsd} are the partial first sentences.
\end{tcolorbox}
\end{center}
\vspace{-0.1cm}
\
\section{Impacts on the Performance of Code Summarization}
\label{sec:modelperformance}
In this section, we investigate the impact of noisy data on the performance of code summarization. Specifically, we choose three state-of-the-art code summarization models and train these models on three versions (\textit{i.e.}, original, controlled, and filtered) of each benchmark dataset.  Thus, we have 3 (the number of models) $\times$ 4 (the number of benchmark datasets) $\times$ 3 (the number of versions per benchmark dataset) = 36 experimental models in total. We evaluate the performance of all the models based on commonly-used metrics for code summarization tasks. 

\subsection{Experimental Design}
\begin{table}[b!]
\centering
\vspace{-0.1in}
\caption{Total and experimental datasets for impact analysis}.
\label{tab:exp_and_total_v2}
\vspace{-0.1in}
\resizebox{\columnwidth}{!}{
\begin{tabular}{|ccc|c|c|c|c|}
\hline
\multicolumn{3}{|c|}{}                                                                                         & \textbf{\funcom}    & \textbf{\tlc}    & \textbf{\codenn}     & \textbf{\pcsd}   \\ \hline
\multicolumn{1}{|c|}{\multirow{2}{*}{\textit{Total}}}        & \multicolumn{2}{c|}{Origin}                              & 2,149,121 & 87,136 & 496,688  & 105,540 \\ \cline{2-7} 
\multicolumn{1}{|c|}{}                              & \multicolumn{2}{c|}{Filtered}        & 1,316,532 & 68,743 & 351,394  & 95,793 \\ \hline \hline
\multicolumn{1}{|c|}{\multirow{4}{*}{\textit{Experimental}}} & \multicolumn{1}{c|}{\multirow{3}{*}{Train}} & Origin     & 1,937,136 & 69,708 & 454,451 & 63,324 \\ \cline{3-7} 
\multicolumn{1}{|c|}{}                              & \multicolumn{1}{c|}{}                       & Controlled & 1,184,438 & 53,597 & 323,226 & 57,849 \\ \cline{3-7} 
\multicolumn{1}{|c|}{}                              & \multicolumn{1}{c|}{}                       & Filtered   & 1,184,438 & 53,597 & 323,226 & 57,849 \\ \cline{2-7} 
\multicolumn{1}{|c|}{}                              & \multicolumn{1}{c|}{Test}                   & Filtered   & 69,392    & 7,584  & 19,319  & 19,028 \\ \hline
\end{tabular}
}
\vspace{-0.3cm}
\end{table}
\begin{table*}[tb]
\centering
% \vspace{-0.1in}
\caption{Performance of existing models trained over different datasets}
\vspace{-0.1in}
\label{table:rq3_result}
\resizebox{0.96\textwidth}{!}{
\begin{tabular}{|c|c|c|c|l|l|l|l|l|l|l|} 
\hline
\textbf{Benchmark} & \textbf{Model}                 & \textbf{Train set}  & \textbf{Training Hours} & \multicolumn{1}{c|}{\textbf{BLEU-1}} & \multicolumn{1}{c|}{\textbf{BLEU-2}} & \multicolumn{1}{c|}{\textbf{BLEU-3}} & \multicolumn{1}{c|}{\textbf{BLEU-4}} & \multicolumn{1}{c|}{\textbf{ROUGE}} & \multicolumn{1}{c|}{\textbf{METEOR}} & \multicolumn{1}{c|}{\textbf{CIDEr}}  \\ 
\hline \hline
\multirow{9}{*}{\funcom}                                      & \multirow{3}{*}{NNGen}     & Origin             &  5h                       &23.87                                     &14.28                                     &11.4                                     &10.05                                     &26.88                                     &12.59                                      &1.26                                      \\
                                                             &                                & Controlled         & 5h                         &21.93                                     &12.19                                     &9.34                                     &8.09                                     &24.84                                     &11.39                                      &1.05                                      \\
                                                             &                                & Filtered           & 5h                         &\textbf{24.58}\ \ \ 3.0\% ↑                                     &\textbf{15.26}\ \ \ 6.9\% ↑                                     &\textbf{12.49}\ \ \ 9.6\% ↑                                     &\textbf{11.2}\ \ \ 10.3\% ↑                                     &\textbf{27.08}\ \ \ 0.7\% ↑                                     &\textbf{13.24}\ \ \ 5.2\% ↑                                      &\textbf{1.38}\ \ \ 9.5\% ↑                                      \\ 
\cline{2-11}
                                                             & \multirow{3}{*}{NCS}       & Origin             & 20h                        & 29.95                               & 17.79                               & 10.2                                & 6.42                                & 34.84                               & 16.14                                & 1.38                                 \\
                                                             &                                & Controlled         &  20h                       & 29.33                               & 17.06                               & 9.78                                & 5.31                                & 34.05                               & 15.65                                & 1.42                                 \\
                                                             &                                & Filtered           &   20h                      & \textbf{30.53}\ \ \ 1.9\% ↑               & \textbf{18.79}\ \ \ 5.6\% ↑                & \textbf{11.47}\ \ \ 12.5\% ↑              & \textbf{7.64}\ \ \ 16.0\% ↑                & \textbf{35.42}\ \ \ 1.7\% ↑                & \textbf{16.32}\ \ \  1.1\% ↑                 & \textbf{1.46}\ \ \ 5.8\% ↑                  \\ 
\cline{2-11}
                                                             & \multirow{3}{*}{Rencos} & Origin             &  9h                       & 27.23                               & 15.97                               & 9.62                                & 6.43                                & 31.97                               & 14.32                                & 1.25                                 \\
                                                             &                                & Controlled         &  9h                & 26.90                               & 15.71                               & 9.48                                & 6.42                                & 31.79                               & 14.16                                & 1.23                                 \\
                                                             &                                & Filtered           & 9h                     & \textbf{27.92}\ \ \  2.5\% ↑                & \textbf{16.8}\ \ \ 5.2\% ↑               & \textbf{10.61}\ \ \ 10.3\% ↑             & \textbf{7.44}\ \ \ 13.6\% ↑                 & \textbf{32.51}\ \ \  1.7\% ↑               & \textbf{14.50}\ \ \ 1.3\% ↑                & \textbf{1.31}\ \ \ 4.8\% ↑                  \\ 
\hline \hline
\multirow{9}{*}{\tlc}                                         & \multirow{3}{*}{NNGen}     & Origin             &  <1h                       & 32.58                               & 24.16                               & 21.92                               & 20.74                               & 36.07                               & 18.14                                & 2.01                                 \\
                                                             &                                & Controlled         &  <1h                & 39.84                               & 32.01                               & 29.24                               & 27.51                               & 43.57                               & 23.22                                & 2.64                                 \\
                                                             &                                & Filtered           &  <1h                       & \textbf{46.88}\ \ \ 43.9\% ↑            & \textbf{39.27}\ \ \  62.5\% ↑               & \textbf{36.81}\ \ \  67.9\% ↑               & \textbf{35.19}\ \ \ 41.1\% ↑               & \textbf{49.08}\ \ \ 36.1\% ↑               & \textbf{25.53}\ \ \ 40.7\% ↑                & \textbf{3.62}\ \ \ 80.5\% ↑                 \\ 
\cline{2-11}
                                                             & \multirow{3}{*}{NCS}       & Origin             &   6h                      & 42.09                               & 32.95                               & 29.09                               & 27.09                               & 46.30                                & 24.18                                & 2.65                                 \\
                                                             &                                & Controlled         &   6h               & 39.28                               & 29.61                               & 25.83                               & 23.89                               & 43.49                               & 22.11                                & 2.37                                 \\
                                                             &                                & Filtered           &   6h                      & \textbf{46.52}\ \ \ 10.5\% ↑               & \textbf{37.19}\ \ \ 12.9\% ↑               & \textbf{33.41}\ \ \  14.9\% ↑               & \textbf{31.38}\ \ \ 13.7\% ↑               & \textbf{49.40}\ \ \ 6.7\% ↑                & \textbf{24.67}\ \ \ 2.0\% ↑                 & \textbf{3.30}\ \ \  24.5\% ↑                 \\ 
\cline{2-11}
                                                             & \multirow{3}{*}{Rencos} & Origin             &  6h                       & 43.66                               & 34.82                               & 31.29                               & 29.19                               & 47.87                               & 24.95                                & 2.81                                 \\
                                                             &                                & Controlled         &  6h                & 43.71                               & 34.89                               & 31.21                               & 28.93                               & 47.85                               & 25.37                                & 2.84                                 \\
                                                             &                                & Filtered           &    6h                     & \textbf{51.54}\ \ \ 18.0\% ↑               & \textbf{42.90}\ \ \ 23.2\% ↑               & \textbf{39.22}\ \ \ 25.3\% ↑               & \textbf{37.00}\ \ \  21.1\% ↑               & \textbf{54.25}\ \ \  13.3\% ↑               & \textbf{28.21}\ \ \ 13.1\% ↑                & \textbf{3.88}\ \ \  38.1\% ↑                 \\ 
\hline \hline
\multirow{9}{*}{\codenn}                                      & \multirow{3}{*}{NNGen}     & Origin             &  <1h                       & 14.86                                & 6.08                                & 4.07                                 & 3.42                                & 18.04                                & 8.54                                 & 0.40                                 \\
                                                             &                                & Controlled         & <1h                 &13.95                                 & 5.09                                & 3.21                                & 2.62                                & 17.08                                 & 7.97                                 & 0.34                                 \\
                                                             &                                & Filtered           &  <1h                       & \textbf{19.89}\ \ \ 33.8\% ↑               & \textbf{8.28}\ \ \ 36.2\% ↑                & \textbf{5.72}\ \ \ 40.5\% ↑               & \textbf{4.96}\ \ \  31.0\% ↑               & \textbf{23.17}\ \ \ 28.4\% ↑                & \textbf{9.67}\ \ \  13.2\% ↑                 & \textbf{0.65}\ \ \ 62.5\% ↑                  \\ 
\cline{2-11}
                                                             & \multirow{3}{*}{NCS}       & Origin             &  15h                       & 25.47                               & 12.34                                & 5.81                                & 3.02                                & 30.47                               & 12.48                                 & 0.80                                 \\
                                                             &                                & Controlled         &  15h                & 25.45                                 & 12.29                                & 5.68                                & 2.88                                 & 31.17                               & 12.30                                 & 0.82                                 \\
                                                             &                                & Filtered           &  15h                       & \textbf{28.68}\ \ \  12.6\% ↑                & \textbf{14.01}\ \ \  13.5\% ↑                & \textbf{6.96}\ \ \  19.8\% ↑                & \textbf{3.87}\ \ \ 22.0\% ↑                & \textbf{34.29}\ \ \ 12.5\% ↑               & \textbf{13.84}\ \ \ 10.9\% ↑                  & \textbf{0.95}\ \ \ 18.8\% ↑                 \\ 
\cline{2-11}
                                                             & \multirow{3}{*}{Rencos} & Origin             & 11h                        & 16.99                                & 7.65                                & 4.09                                & 2.64                                & 20.91                               & 8.33                                 & 0.49                                 \\
                                                             &                                & Controlled         &    11h                     & 16.30                                 & 7.09                                & 3.75                                 & 2.43                                 & 20.00                               & 8.13                                 & 0.44                                  \\
                                                             &                                & Filtered           &  11h                       & \textbf{24.72}\ \ \ 45.5\% ↑                 & \textbf{11.36}\ \ \ 48.5\% ↑                & \textbf{6.51}\ \ \ 59.2\% ↑                 & \textbf{4.56}\ \ \ 42.1\% ↑              & \textbf{29.35}\ \ \ 40.4\% ↑               & \textbf{11.52}\ \ \ 38.3\% ↑                 & \textbf{0.82}\ \ \ 67.3\% ↑                 \\ 
\hline \hline
\multirow{9}{*}{\pcsd}                                        & \multirow{3}{*}{NNGen}     & Origin             &  <1h                       & 22.52                               & 15.48                               & 12.63                               & 10.45                               & 24.90                                & 12.97                                & 1.24                                 \\
                                                             &                                & Controlled         & <1h                        & 21.81                               & 14.77                               & 11.99                               & 9.91                                & 24.16                               & 12.49                                & 1.18                                 \\
                                                             &                                & Filtered           & <1h                        & \textbf{25.96}\ \ \ 15.3\% ↑               & \textbf{18.91}\ \ \ 22.2\% ↑               & \textbf{16.27}\ \ \ 28.8\% ↑               & \textbf{14.00}\ \ \ 25.4\% ↑               & \textbf{27.68}\ \ \ 11.2\% ↑               & \textbf{15.09}\ \ \ 16.3\% ↑                & \textbf{1.63}\ \ \ 31.9\% ↑                 \\ 
\cline{2-11}
                                                             & \multirow{3}{*}{NCS}       & Origin             &    6h                     & 28.14                               & 18.69                               & 14.28                               & 11.36                               & 32.95                               & 16.30                                 & 1.61                                 \\
                                                             &                                & Controlled         & 6h                        & 26.85                               & 17.42                               & 13.05                               & 10.17                               & 31.77                               & 15.42                                & 1.49                                 \\
                                                             &                                & Filtered           &  6h                       & \textbf{37.33}\ \ \ 32.7\% ↑               & \textbf{24.74}\ \ \ 32.4\% ↑               & \textbf{19.49}\ \ \ 36.5\% ↑               & \textbf{16.48}\ \ \ 31.1\% ↑               & \textbf{40.93}\ \ \ 24.2\% ↑               & \textbf{18.67}\ \ \ 14.5\% ↑                & \textbf{2.07}\ \ \ 28.6\% ↑                 \\ 
\cline{2-11}
                                                             & \multirow{3}{*}{Rencos} & Origin             & 5h                 & 30.37                               & 21.27                               & 16.42                               & 12.93                               & 33.66                               & 17.40                                 & 1.65                                 \\
                                                             &                                & Controlled         &  5h                       & 29.73                               & 20.55                               & 15.71                               & 12.37                               & 33.05                               & 16.96                                & 1.59                                 \\
                                                             &                                & Filtered           &    5h                     & \textbf{33.59}\ \ \ 10.6\% ↑               & \textbf{24.14}\ \ \ 13.5\% ↑               & \textbf{19.63}\ \ \ 19.5\% ↑               & \textbf{16.10}\ \ \ 19.7\% ↑               & \textbf{36.15}\ \ \ 7.4\% ↑                & \textbf{19.18}\ \ \ 10.2\% ↑                & \textbf{2.02}\ \ \ 22.4\% ↑                 \\
\hline
\end{tabular}
}
\vspace{-0.1cm}
\end{table*}

\subsubsection{Data Preparation}
We use three versions of each benchmark dataset as training sets, as shown in Table \ref{tab:exp_and_total_v2}. 
The ``Total'' rows illustrate the overall data before and after being distilled by our tool.
The ``Experimental'' rows show the data that are used for our experiments.
The ``Origin" refers to the original training dataset split by the benchmark dataset. 
The ``Filtered" refers to the train/test dataset cleaned by our tool from the ``Origin" train/test datasets. To benchmark the performance variation brought by the size shrinking, we further build 
the ``Controlled" set by randomly sampling from the ``Origin" set, which has an equal amount of data instances as the ``Filtered" dataset.

\subsubsection{Code Summarization Models} {As introduced in Section~\ref{sec:codeSum}, existing code summarization models can be divided into three categories}: Information Retrieval (IR) based approaches, Neural Machine Translation (NMT) based approaches, and hybrid approaches that combine IR and NMT techniques. We select one state-of-the-art method from each category to explore the impact of noisy data on model performance.

\textbf{NNGen~\cite{liu2018neural}} is an IR-based model for generating commit messages by utilizing the nearest neighbors algorithm. It first embeds code changes into vectors based on the bag of words and the term frequency. Then, NNGen retrieves the nearest neighbors of code changes by calculating the cosine similarity of vectors and the BLEU-4 score. Finally, it directly chooses the message of the code change with the highest BLEU score as the final result. 
% \footnote{https://github.com/Tbabm/nngen}

\textbf{NCS~\cite{ahmad2020transformer}} is an NMT-based model which replaces the previous RNN units with the more advanced Transformer~\cite{vaswani2017attention} model. NCS extends the vanilla Transformer in two aspects. Firstly, it incorporates the copying mechanism~\cite{see2017get} in the Transformer to allow both generating words from vocabulary and copying from the input source code. Secondly, NCS utilizes relative positional embedding rather than absolute positional embedding to capture the semantic representation of the code better. 
% \footnote{https://github.com/wasiahmad/NeuralCodeSum}

\textbf{Rencos~\cite{zhang2020retrieval}} is a state-of-the-art model that combines the advantages of both IR-based and NMT-based techniques. Specifically, given an input code snippet, Rencos first retrieves its two most similar code snippets in the training set from the aspects of syntax and semantics, respectively. Then it encodes the input and two retrieved code-snippets, and generates the summary by fusing them during decoding.
% \footnote{https://github.com/zhangj111/rencos}

\subsubsection{Experimental Settings.}
% When training the three code summarization models with benchmark datasets, we use the implementation provided by the original paper. We adopt the recommended hyperparameter settings. Specifically, when training NNGen model, we set the number of candidate neighbors to 5 for all benchmark datasets. When training NCS model, we set the \textit{learning\_rate}=1e-4. For {\tlc}, {\pcsd}, and {\codenn}, we set \textit{batch\_size}=32, \textit{max\_epoch}=50. For {\funcom}, we set \textit{batch\_size}=128, and \textit{training\_epoch}=20. When training Recos model, we set the \textit{learning\_rate} = 1e-3, \textit{batch\_size}=32, \textit{max\_iteration}=100k. For a fair comparison, all experiments use the greed search strategy to generate summaries. Note that, XXX. \lin{@fangwen, explain sth like the performance will slight lower than reported, and we do not achieve best performance, we focus on increment}
The experimental environment is a desktop computer equipped with an NVIDIA GeForce RTX 3060 GPU, Intel Core i5 CPU, 12GB RAM, running on Ubuntu OS. When training the three code summarization models with the benchmark datasets, we follow the implementation provided in their original papers, and adopt the recommended hyperparameter settings, except for the training epoch of NCS and Rencos. To save training time and computation resources, we set \textit{max\_epoch}=50 for NCS and \textit{max\_iteration}=100k for Rencos. 
%Lowing the training epoch may result in suboptimal performance of the model than the performance reported in the literature. However, the evaluation conclusion on the effectiveness of our cleaning tool should not be affected, which is obtained by comparing the performance changes of the models before and after noise removal, rather than the absolute value of the model performance. \lin{@wangsong}

\subsubsection{Evaluation Metrics}
We evaluate the performance of the three models using four metrics including BLEU~\cite{DBLP:conf/acl/PapineniRWZ02}, METEOR~\cite{DBLP:conf/acl/BanerjeeL05}, ROUGE-L~\cite{lin2004rouge}, and CIDEr~\cite{DBLP:conf/cvpr/VedantamZP15}.
\textbf{BLEU} measures the $n$-gram precision by computing the overlap ratios of $n$-grams and applying brevity penalty on short translation hypotheses. 
BLEU-1/2/3/4 correspond to the scores of unigram, 2-grams, 3-grams, and 4-grams, respectively. 
\textbf{ROUGE-L} is defined as the length of the longest common subsequence between generated sentence and reference, and based on recall scores.
\textbf{METEOR} is based on the harmonic mean of unigram precision and recall, with recall weighted higher than precision. 
\textbf{CIDEr} considers the frequency of $n$-grams in the reference sentences by computing the TF-IDF weighting for each $n$-gram. ${CIDEr}_n$ score for $n$-gram is computed using the average cosine similarity between the candidate sentence and the reference. 

% \textbf{BLEU} is a standard evaluation metric in the source code summarization works. BLEU score calculates the geometric mean of $n$-gram matching precision scores, which is multiplied by a brevity penalty to prevent very short generated sentence. BLEU-1/2/3/4 correspond to the scores of unigram, 2-grams, 3-grams and 4-grams. 
% \textbf{ROUGE-L} is widely used in text summarization tasks. The score allows multiple references since there may be multiple correct summaries of a text. The metric is defined as the length of the longest common subsequence between generated sentence and reference, and based on recall scores.
% \textbf{METEOR} is used primarily in machine translation tasks in the NLP literature. The metric is based on the harmonic mean of unigram precision and recall, with recall weighted higher than precision. METEOR can produce a good correlation with human judgment at the sentence or segment level.
% \textbf{CIDEr} is usually used for measuring the quality of image captions, which considers the frequency of $n$-grams in the reference sentences by computing the TF-IDF weighting for each $n$-gram. ${CIDEr}_n$ score for $n$-gram is computed using the average cosine similarity between the candidate sentence and the reference. 

\subsection{Quantitative Results}
Table~\ref{table:rq3_result} shows the performance of the three models trained over different experimental datasets. 
Overall, removing noisy data {from the training set} in the four benchmark datasets produces a positive effect on improving the performance of the three models. {Training three existing models with the filtered benchmark datasets improves the BLEU-4
by 26.9\%, 20.7\%, and 24.1\%, ROUGE by 19.1\%, 11.3\%, and 15.7\%, METEOR by
18.9\%, 7.1\%, and 15.7\%, CIDEr by 46.1\%, 19.4\%, and 33.2\%, respectively.}
%\rev{The average increases of the three models are 6.9\%, 33.2\%, and 24.7\% on all seven metrics, respectively.}%\song{it's inappropriate to average different metrics, maybe say BLUE, rouge, meteor, cider separately?}

Among the four benchmark datasets, the effect on the {\csn} dataset is the most significant, %leads to the largest increase of the three models, 
which leads to the three models (NNGen, NCS, and Rencos) increasing by 31.0\%, 22.0\% and 42.1\% on BLEU-4, 28.4\%, 12.5\% and 40.4\% on ROUGE, 13.2\%, 10.9\%, and 38.3\% on METEOR, and 62.5\%, 18.8\%, and 67.3\% on CIDEr, respectively. This is followed by {\tlc} and {\pcsd}. Considering the fact that even the least effect obtained in {\funcom} still contributes to an increase of 10.3\%, 16.0\% and 13.6\% on BLEU-4, 0.7\%, 1.7\% and 1.7\% on ROUGE, 5.2\%, 1.1\%, and 1.3\% on METEOR, and 9.5\%, 5.8\%, and 4.8\% on CIDEr, respectively. %where NNGen increases 7.9\%, NCS increases 8.4\%, and Rencos increase 4.6\% on average.) 
%Based on the above results, 
{The main reason that the three models exert the biggest performance difference on CSN is that, the primary noisy data on CSN are content tampering by HTML tags, and removing these noises will make the generated comments more accurate. We will illustrate this in the following qualitative analysis.}
We argue that the existing models used for code summarization tasks in the literature have a significant scope of improvement given a large, good-quality dataset.

By observing the performance of the three models trained on different filtered datasets, we find that the relative ranking among the three types of models is not consistent. For the filtered {\tlc}, Rencos achieves the best performance on all metrics compared to the other two models. While NCS performs best when trained on filtered {\funcom}, {\csn}, and {\pcsd}. This result implies that, to more comprehensively evaluate different code summarization models, it is better to use multiple datasets, as the ranking of the model can be inconsistent on different datasets.

% %Diversity of evaluation datasets.
% When analyzing the impacts of noisy data on code-summarization model performance, we also notice that, the ranking among the three types of models does not preserve when evaluating them on different datasets. For example, Rencos (IR+NN) outperforms other models in TLC but is worse than NSC (NN) in Funcom and PCSD, as shown in Table \ref{table:rq3_result}. 
% To more comprehensively evaluate different code-summarization models, it is recommended to use multiple datasets, as the ranking of the model can be inconsistent on different datasets.

\begin{center}
\begin{tcolorbox}[colback=gray!10,%gray background
                  colframe=black,% black frame colour
                  width=8.3cm,% Use 8cm total width,
                  arc=1mm, auto outer arc,
                  boxrule=0.5pt,
                  left=3pt,
                  right=3pt,
                  top=3pt,
                  bottom=3pt
                 ]
\textbf{Finding 2}: Removing noisy data {from the training set} in the four benchmark datasets has a positive influence on the performance of the models.
{Training three existing models with the filtered benchmark datasets improves the BLEU-4 by 26.9\%, 20.7\%, and 24.1\%, respectively.}
%ROUGE by 19.1\%, 11.3\%, and 15.7\%, METEOR by 18.9\%, 7.1\%, and 15.7\%, CIDEr by 46.1\%, 19.4\%, and 33.2\%,
%\rev{
%with an average increment of 6.9\%, 33.2\%, and 24.7\%, respectively.}%\song{missing metrics here}
\end{tcolorbox}
\end{center}

\subsection{Qualitative Analysis}
\label{sec:quaAna}
\begin{figure}[tbp]
% \vspace{-0.1cm}
\centering
\subfigure[Case 1: An example of  generated comments that contains over-splitting variable, HTML tags, and unfinished sentence]{
    \label{fig:example-2}
    \includegraphics[width=0.95\columnwidth]{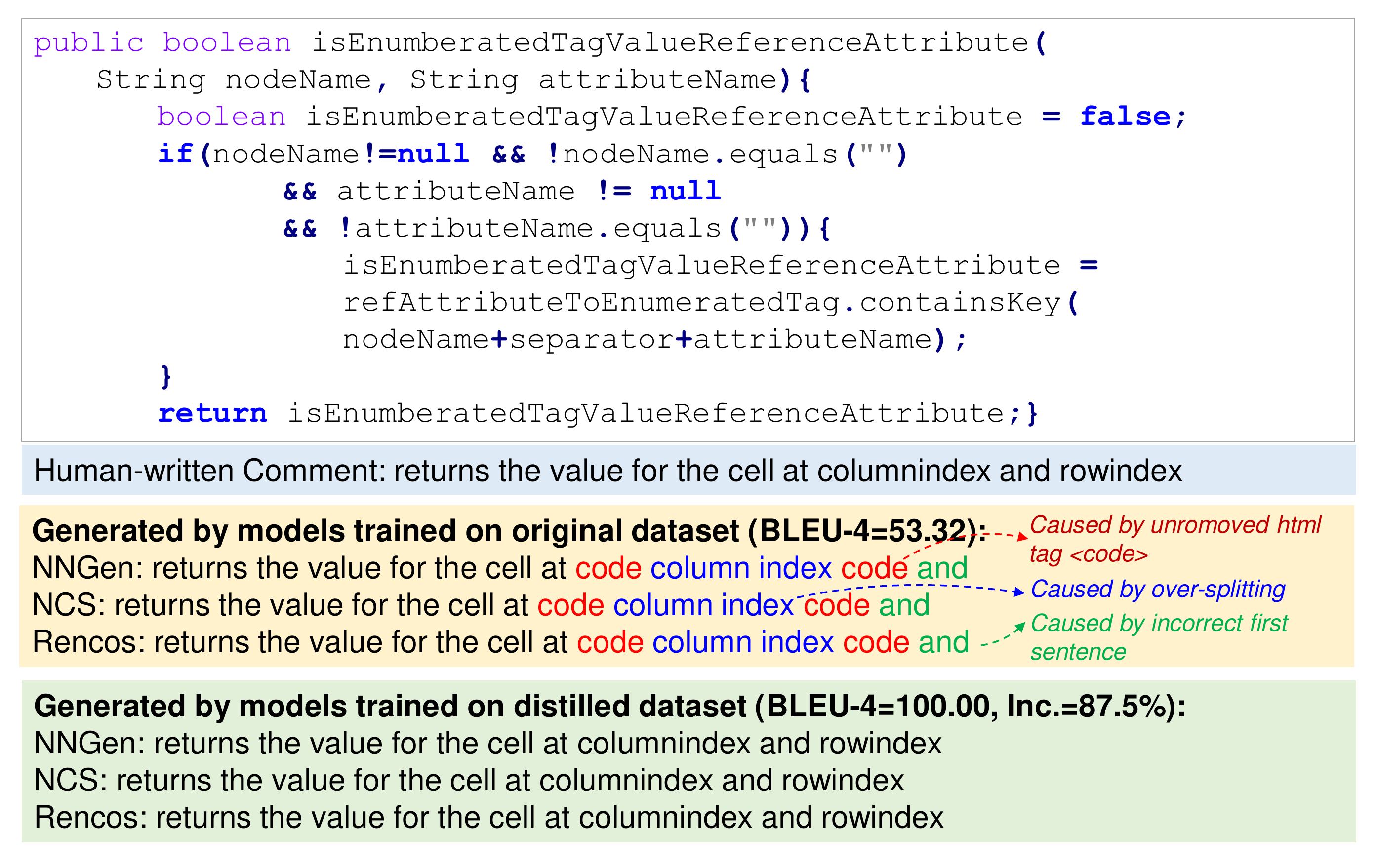}

}
\subfigure[Case 2: An example of  generated comments that contains HTML tags]{
    \label{fig:example-1}
    \includegraphics[width=0.95\columnwidth]{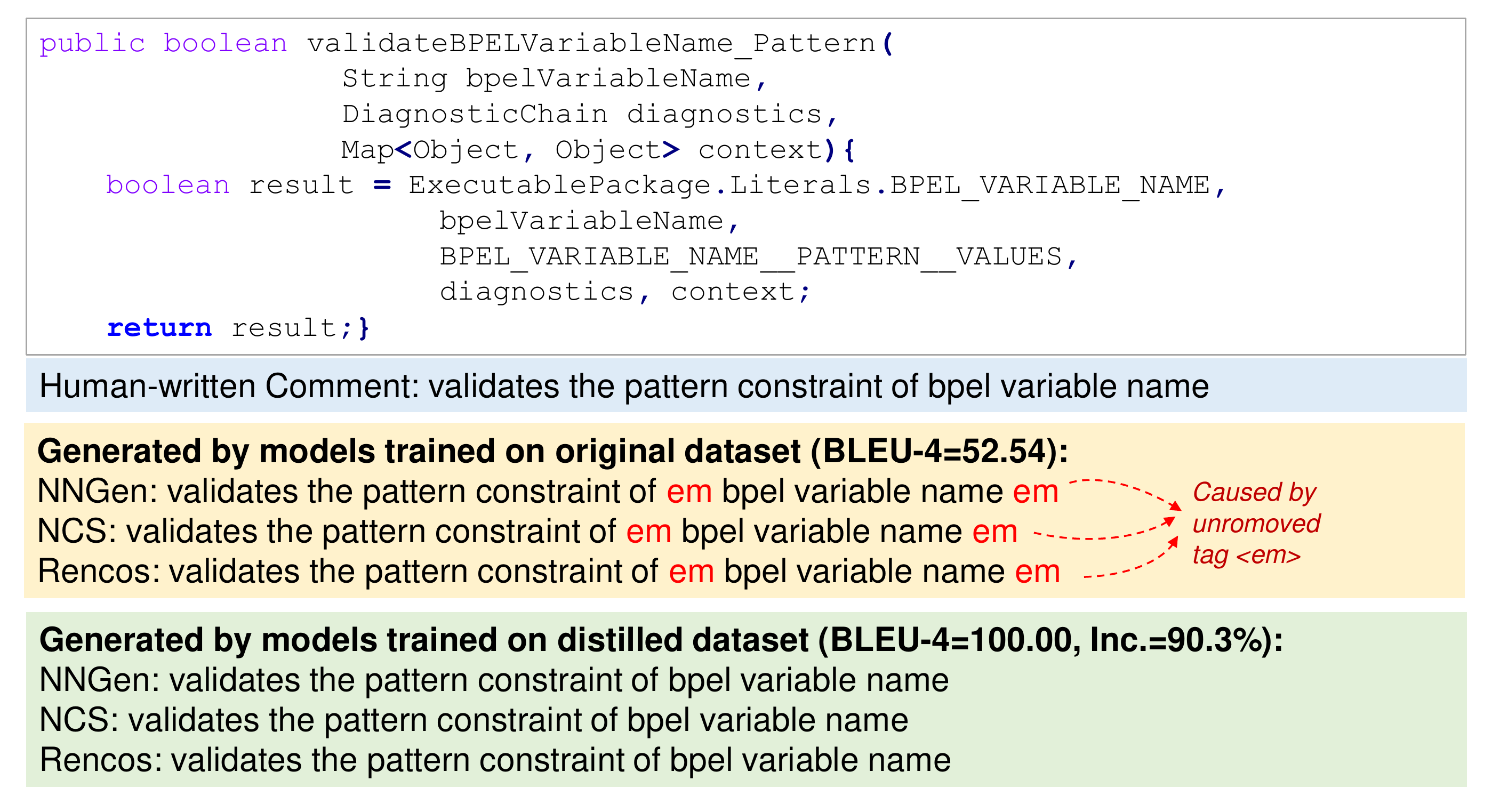}
}
\vspace{-0.4cm}
\caption{Inaccurate comment generation affected by noises}
\label{fig:cases}
\vspace{-0.6cm}
\end{figure}

To qualitatively illustrate the impact of the noises on code summarization models, we present two cases generated by the three models trained on different datasets, as shown in Figure~\ref{fig:cases}. %\ref{tab:rq3_example}. 
\textbf{Overall}, the comments generated by the models trained on the distilled datasets tend to be more accurate and more readable than the comments generated by the models trained on the origin datasets.

\textbf{Case 1.} 
Given the code, the comments generated by the three models trained on origin benchmark datasets are ``returns the value for the cell at code column index code and''. Compared with the human-written comment, we consider the following three errors are likely related to noisy data:
(1) The over-splitting ``columnindex'' as ``column index''. This error is likely to be caused by the over-splitting of variable identifiers in the comment;
(2) The redundant ``code'' around ``\textit{columnindex}''. It is mainly due to the fact that the origin datasets contain many unremoved HTML tags, thereby increasing the probability of HTML tags, \textit{e.g.}, ``<code>'', appearing in the context, making it easier for the model to generate these HTML words when encountering some specific patterns;
(3) The redundant ``and'' in the end. This is mainly because the widely-existing partial or verbose sentences in training sets would hinder the ability of the model to learn to determine when the generation process should end.
After being retrained on the distilled data, the three models can accurately generate the comments, where the BLEU-4 has an 87.5\% increase, from 53.32 to 100.00. 

\textbf{Case 2.} Given the code, the comments generated by the three models trained on origin benchmark datasets have two redundant ``em'', which are caused by the unremoved HTML tag ``<em>'' that is
used to define emphasized text.
After being retrained on the distilled data, the three models can accurately generate the comments, where the BLEU-4 has a 90.3\% increase, from 52.54 to 100.00.

\section{Discussion}
\label{sec:threats}
In this section, we discuss several interesting implications that are derived from the results of this study, aiming to facilitate the code summarization research and the SE community.
%serve as the practical guidelines for improving the performance of automated code summarization regarding the collection and selection of benchmark data.

\subsection{Impact of Noises on Code Summarization Datasets and Models}
\subsubsection{Impact of Noises on Datasets.}
\textbf{(1) Removing the noises in Funcom leads to a slight improvement in model performance} (i.e., the BLEU-1 score increases 2.46\% on average). It might be because that Funcom is the one with the most auto-generated code, and auto code offers “easy gain” in performance that is not available anymore. Therefore, the baseline performance could actually decrease if removing them from testset, thus making the improvements of other rules look smaller in comparison.\textbf{ (2) Removing the noises in TLC, CSN, and PCSD leads to a considerable improvement in performance} (i.e., the BLEU-1 score increases 24.1\%, 30.6\%, and 19.5\% on average respectively. It might be because the major noises of these three datasets are `Verbose Sentence', `Content Tampering', and `Partial Sentence', respectively, and removing them will benefit the models. 

\subsubsection{Impact of Noises on Models.}
Based on the results shown in Table 4 and Table 6, we can observe that the noises affect code summarization models differently. 
\textbf{(1) Removing `Verbose Sentence' noises might largely benefit the IR-based model NNGen.} The major noises in TLC are `Verbose Sentence', and removing them leads to the performance of NNGen increases 53.25\% on average of all the seven metrics, followed by Rencos (21.74\%) and NCS (12.17\%). It might be because the NNGen model directly outputs the retrieved results. Taking the retrieved verbose comments as the output leads to a substantial decline in the scores of the evaluation metrics, since these N-gram based metrics are less beneficial for longer comments.
\textbf{(2) Removing `Content Tampering' noises might largely benefit the hybrid model Rencos.} The major noises in CSN are `Content Tampering', and removing them leads to the performance of Rencos increasing 48.75\% on average, followed by NNGen (35.11\%) and NCS (15.73\%). 
It might be because the hybrid model Rencos employs a more complex input including the test code and the retrieved data. Such models' effective training typically requires the ground-truth comments to be semantically correct. However, the `Content Tampering' noises cause the ground-truth comments to mingle with irrelevant text such as HTML tags or URLs, which alter the semantics of the ground-truth comments. Therefore, when removing the `Content Tampering' noises, the hybrid model Rencos increases the most.
\textbf{(3) Removing `Partial Sentence' noises might largely benefit the NMT-based model NCS.} The major noises in PCSD are `Partial Sentence', and removing them leads to the performance of NCS increasing 28.56\% on average, followed by NNGen (21.58\%) and Rencos (14.77\%). The main reason is that the comments belonging to the `Partial Sentence' noise are not complete sentences, and thus lack integrity for both syntactic and semantic, which hinders language models like NCS from learning the syntactic and semantic information correctly. 
We will further illustrate the impact of noises in qualitative cases.

\subsection{Lessons Learned of Data Preprocessing for Code Summarization}

Data quality has been a growing concern, especially since deep learning (DL) is widely applied for massive SE tasks. As DL models typically require high-volume data, ensuring data quality at a large scale has become a compelling need. %more challenging. 
Most existing studies focus on advances in modeling but typically overshadow the data quality. When investigating the current code summarization models (as shown in Table \ref{table:dataset_intro}), we notice that a substantial amount of data preprocessing operations are cursory or lack consistency. 
In addition, there is a lack of principles or methods to guide and reinforce the data preprocessing associated effort while conducting code summarization research, regarding how to soundly preprocess benchmark datasets for different tasks.
For instance, our results show that noisy data extensively exist in the four widely-used benchmark datasets for code summarization tasks. Shi \textit{et al.}~\cite{DBLP:journals/corr/abs-2107-07112} reported that different code preprocessing operations can affect the overall performance of code summarization models by a noticeable margin. 
Therefore, paying more attention to data quality while training code summarization models is recommended, rather than directly reusing existing datasets without quality inspection. 

Specifically, the study and its results lead to the identification of the following lessons learned from the quality assessment on benchmark datasets, for future code summarization researchers.
%to build their own datasets. 
%[leftmargin=4mm]
\begin{itemize}
\item When reusing existing datasets, check the quality of processed data by comparing with their raw data. 
\item Extracting the first sentences of comments is error-prone.
\item Avoid including under-development or obsolete code, \textit{e.g.}, TODOs, commented-out methods.
\item Avoid over-splitting of variable identifiers in comments.
\item Be careful of “what to comment”, check whether the following types of code-comment data are suitable for your scenarios: interrogative comments, auto code, duplicated code, and block-comment code.
\item Remember to deal with abnormality, \textit{e.g.}, HTML tags, URL, code path, and non-literal natural languages.

% \item Assuming the first sentences of comments is representative. Rationale: the results show that using first sentences of comments is error-prone. 
% \item More comment is better. Rationale: the results show that we should be careful of ``what to comment'', check whether the following types of code-comment data are suitable for your scenarios: interrogative comments, auto code, duplicated code, block-comment code.
% \item Over-splitting of variable identifiers in comments. Rationale: ...
% \item Including under-development or out-of-date code, e.g., TODOs, commented-out methods. Rationale: ...
% \item Remember to deal with abnormality, e.g., HTML tags, URL, code path, non-literal natural languages. \ye{This sentence needs to be converted into the description of an anti-pattern.}
\end{itemize}

\subsection{Tool Support and Potential Applications}

We release the implementation of the {\tool} code-comment cleaning tool as a third-party Python library~\cite{cat_pypi}, which can be easily integrated into the development pipeline of most code summarization models. The features of {\tool} are:
(1) \textbf{Configurable and Extendable Rules}. The ruleset in {\tool} is configurable, which allows users to customize the existing rules based on the different data characteristics. Besides, {\tool} provides interfaces enabling users to design new rules or clean functions to extend the feature of {\tool}.
(2) \textbf{Support for Multiple Programming Language}. {\tool} is a tool that can support multiple programming languages such as Java, Python, and C\#. Similarly, for applying {\tool} to other programming language datasets, users can replace the existing language-specific rules with the new rules.
We also release the distilled four benchmark datasets on our website~\cite{website} to facilitate future code summarization research.

This study applies the {\tool} tool for exploring the data quality issues for code summarization exclusively. 
Similar to code summarization, there exist other tasks on the intersection of Natural Language Processing and Software Engineering, such as {commit message generation~\cite{jiang2017automatically} (generates a natural language summarization for each submitted code change), code search~\cite{gu2016deep} (generates API usage sequences for a given natural language
query), and code synthesis~\cite{wei2019code} (synthesizes code based on natural language intents).
These tasks also require datasets that contain a large number of code and natural language description pairs to train their models, the quality of their datasets can have a critical impact on the performance of models. Thus, we recommend future research on these topics should also apply {\tool} to remove potential data preprocessing noises.}
In addition, we believe {\tool} can also facilitate the downstream tasks with building large pre-trained {code models~\cite{karmakar2021pre} to 
learn code representations, 
which require a large corpus of code. {\tool} can help remove code noises as listed in Table~\ref{tab:taxonomy}.}

%\subsection{Implications for improving data quality}
\subsection{Implications for Research and Practice}

%这一部分要强调collaborative community effort，从数据预处理的设计和质量监控、以及论文审稿checklist等等方面给出具体操作建议

\textbf{Need for collaborative community effort on principles of data preprocessing.}
Improving data quality takes collaborative, community effort to establish and maintain principles, methods, and tools to govern the data extraction and preprocessing pipelines. This study takes an initial step towards addressing the challenges of building high-quality datasets for code summarization exclusively. Although we have proposed quality criteria to measure data quality, methods to help collect data, and filters to remove preprocessing noises, our solutions might be not sufficient for other research tasks in software engineering or data science areas that require massive data as inputs because of the diversity of data sources, the complexity of different data structures, and the scale of data volume. Thus, we urge for collaboration and effort from the whole research community to help build a comprehensive and reliable principle set for data collection, preprocessing, and quality assessment, which we believe can benefit our research community.
%\song{updated, feel free to revise}

\textbf{Need for research on comprehensive noisy code-comment detection.}
In this study, we define 12 categories of noisy data in code summarization datasets, and apply our cleaning tool to filter them out. For the filtered data, we observed several additional quality issues that require a deeper understanding of the content of the comments and the corresponding code. 
\textbf{(1) Inconsistent code-comment pairs.}
%in that comments are not consistent with the code in benchmark datasets, \textit{e.g.}, obsolete comments during software evolution.
The following example shows a spotted inconsistency between the comment and its code. The comment explicitly states the return fact, but the code does not.   
Since only a few approaches are proposed 
\begin{lstlisting}
/* Read information object and return pointer */
public void readInformationObject(...){
  try {objectDecoder.checkResolved(infoObj);
  } catch (  final Exception e) {
    LogWriter.writeLog("Exception: " + e.getMessage());
  ...
 \end{lstlisting}
    \vspace{-2mm}
\begin{small}
\texttt{Comment (TLC): read information object and \textcolor{red}{return pointer}}\\
\texttt{Code (TLC): public \textcolor{red}{void} read information object...}
\end{small}
\vspace{2.5mm}

\noindent
for detecting inconsistent Java code-comment pairs  \cite{DBLP:journals/sqj/CorazzaMS18, DBLP:conf/iwpc/RabbiS20}, and there is a lack of inconsistent code-comment detection for other programming languages, such as Python and C\#, it is quite challenging to assess and clean such noises in a parallel corpus. 
\textbf{(2) Low-readability comments}. We noted that some comments are not fluency or have syntactic errors. E.g., a comment in the PCSD dataset ``transforms a doc in content in one document in presentation''. 
If trained on datasets with such comments, the end-to-end code summarization models are also at risk of producing low-readability comments.
\textbf{(3) Less-informative comments}. We found that many methods in benchmark datasets do not need comments as methods are self-explainable with their names. For example, two methods in Funcom dataset are named "renderImgTag" and "createTextPane", and their corresponding comments are "render img tag" or "create the text pane". Since such comments are highly similar to the method names, they can hardly convey additional information for better understanding of the source code. If trained on datasets with such comments, the code summarization models are also 
likely to produce less-informative comments.
Therefore, there is a need for research on comprehensive noisy code-comment detection, which can further benefit the quality improvement of code-comment datasets.

%[] = https://github.com/acmsigsoft/open-science-policies
\textbf{Integrating tool support to aid publication peer review.}
Following the open science policies~\cite{open-science-policies}, most existing research makes their raw and transformed data publicly accessible during the peer review process. For those datasets that are in the format of code-comment pairs, the {\tool} cleaning tool can be used to automatically detect their noise data in a reasonable time. The output statistic could objectively reflect the inside quality of the open datasets to some extent, thus can help professional peer reviewers to infer the quality and reliability of the under-review research.

% \textbf{Considerable noise of duplicated code.}
% Despite the debate on the harmfulness of code clone (i.e., duplicated code) in a software project, i.e., on the one hand-code cloning is considered a harmful practice for software maintenance as it requires consistent changes of the entities that share the same cloned fragment~\cite{lozano2007evaluating}, on the other hand, code cloning can often be used in a positive fashion such as maintaining system stability and tracking code ownership~\cite{kapser2008cloning}. 
% In this work, we consider duplicated code as noise and can be harmful to code summarization models. First of all, it is inline with the data preprocessing principle of learning-based models, i.e., the training set should not contain samples in the test set. 
% Second, keeping code duplication can result in model performance and generalizability not being comprehensively evaluated when applied the models to different benchmark datasets that do not share duplicates.\ye{not following the previous sentence.} Therefore, it is considerable to define duplicated code as noisy data. 
% To facilitate other researches that treat them as normal data, we make the duplicate-code rule switchable in our cleaning tool. \ye{See comment.} %Ye: this part is a little distracting...Maybe i didn't get the point.

% %这个有点纠结，因为，如果重复是数据集的客观现象，那就不算是问题。如果重复与数据集的客观分布不同，那就是问题。还有就是，从我的角度，我会认为：如何避免overfitting应该是方法本身应该具备的特性。而不是数据集“应该”帮助算法完成的。
\subsection{Threats to Validity}
%threat: there still may unfiltered noise, hard to find, e.g., inconsistent code and comments. 

One threat to validity relates to the random sampling process. Sampling may lead to incomplete results, \textit{e.g.}, noise taxonomy, we plan to enlarge the analyzed dataset and inspect whether new types of noises are emerging. Moreover, our heuristic rules for data cleaning are elaborated from the four popular benchmark datasets, covering Java and Python. Although we generally believe all similar code-comment datasets may benefit from our cleaning tool, future studies are needed to focus on datasets with other programming languages. 

The second threat might come from the process of manual annotation and card sorting. We understand that such a process is subject to introducing mistakes. To reduce that threat, we establish a labeling team, and perform multiple rounds of labeling to make sure that all participants achieve conceptual coherence about noisy categories.

The third threat relates to the BLEU that is used to evaluate the performance of code summarization models. Recent researchers have raised concern over the use of BLEU \cite{DBLP:conf/sigsoft/RoyFA21}, warning the community that the way BLEU is used and interpreted can greatly affect its reliability. To mitigate that threat, we also adopt other metrics, \textit{i.e.}, ROUGE, METEOR, and CIDEr, when evaluating performance.

Another threat to validity is the replication of each model. To ensure that the experimental results are consistent with their papers, we retrain the models using the source code provided by the authors, and reuse the parameters provided by the authors.  
Our experiments show that the performance of our retrained models is comparable to the performance of models reported in the papers. %Their performances are comparable. 
For example, the METEOR score of Rencos is 21.1 on {\pcsd} in their paper, and our retrained Rencos model is 20.2. 
%The minor difference is inevitable since we evaluate on the filtered test set while they evaluate on the origin test set.  

%Our heuristic rules for data cleaning are elaborated and trained from the four popular benchmark datasets, covering Java and Python. Although we generally believe all similar code-comment datasets may benefit from our cleaning tool, future studies are needed to focus on datasets with other programming languages. 

% However, the evaluation conclusion on the effectiveness of our cleaning tool should not be affected, which is obtained by comparing the performance changes of the models before and after noise removal, rather than the absolute value of the model performance.

\section{Related Work}
\label{sec:relatedwork}
Recently, more and more researchers have realized that there are some underlying threats to the validity of existing code summarization research. These empirical studies mainly focused on data, evaluation metrics, and model effectiveness.

\textbf{Biases in data.}
{Existing research of data biases in the code summarization related tasks mainly focused on data quality, data representativeness, code preprocessing, and data selection.}
%\ye{Reorganize according to major types of data bias such as: data selection, data quality, data representativeness, data pre-processing assumption, etc.}
%data quality
Sun \textit{et al.} \cite{DBLP:journals/corr/abs-2202-06649} applied syntactic and semantic query cleaning to improve the data quality for code search tasks. Their experiment results show that, training the popular code-search model with the filtered dataset improves its performance significantly.
%data representativeness
Gro \textit{et al.} \cite{DBLP:conf/kbse/GrosSDY20} examined the underlying assumption {about data representativeness that:}  the task of generating comments sufficiently resembles the task of translating between natural languages, and so similar models and evaluation metrics could be used. By comparing four code-comment datasets, \textit{i.e.}, CodeNN, DeepCom, FunCom, and DocString, with a standard natural language translator dataset WMT19, they reported that comments are far more saturated with repeating trigrams than English translation datasets, and the repetitiveness has a very strong effect on measured performance.
%data preprocessing
Shi \textit{et al.} \cite{DBLP:journals/corr/abs-2107-07112} analyzed the influence of code preprocessing operations and dataset size on code summarization model performance. They found that different code preprocessing operations can affect the overall performance by a noticeable margin, and the code summarization approaches perform inconsistently on different datasets.
%code commenting necessity
Huang \textit{et al.} \cite{DBLP:journals/spe/HuangJSHCZ20}
{reported the biases in data selection that, not all code is necessarily commented.}
They analyzed 136 well-known projects in GitHub, and reported that only a small part (4.4\%) of methods have header comments in real software projects. They proposed a machine learning technique to automatically identify commenting necessity, based on the structural features, syntactic features, and textual features of code.
There is a lack of in-depth analysis of the benchmark datasets. Our study bridges that gap with a large-scale analysis of {\RQone} and {\RQtwo} in the benchmark datasets, and investigates performance variation of existing models on the distilled dataset.

\textbf{Biases in evaluation metrics. }
Roy \textit{et al.} \cite{DBLP:conf/sigsoft/RoyFA21} conducted an empirical study with 226 human annotators to assess the degree to which automatic metrics reflect human evaluation for code summarization tasks. Their results indicated that metric improvements of less than 2 points do not guarantee systematic improvements in summarization quality, and are unreliable as proxies of human evaluation. 
Gro \textit{et al.} \cite{DBLP:conf/kbse/GrosSDY20} measured 5,000 code-comment pairs, and found that the variants of BLEU chosen can cause substantial variation in the measured performance.
Shi \textit{et al.} \cite{DBLP:journals/corr/abs-2107-07112} also examined the BLEU variants. They concluded that BLEU variants used in prior work on code summarization are different from each other and the differences can carry some risks such as the validity of their claimed results. 
Mahmud \textit{et al.} \cite{DBLP:journals/corr/abs-2106-08415} observed that some auto-generated comments provide a semantic meaning similar to the ground truth, despite exhibiting fewer n-gram matches. Therefore, they argued the feasibility of n-gram metrics such as BLEU. 
Most of these work focus on validating the evaluation procedure for code summarization, while our work targets to validate the benchmark datasets, which would be important and valuable for building sound code summarization models.

%and suggested a modification to the evaluation procedure for code summarization models. 

\textbf{Analysis on Model Effectiveness.}
Mahmud \textit{et al.} \cite{DBLP:journals/corr/abs-2106-08415} compared three recently proposed code summarization models, and performed a manual open-coding of the most common errors committed by the models. They reported that missing information and incorrect construction are the most prevalent error types.
Chen \textit{et al.} \cite{DBLP:journals/tosem/ChenXHLL21} classified code comments into six categories (``what'', ``why'', ``how-to-use'', ``how-it-is-done'', ``property'', and ``others'') according to the intention, and conducted an experiment to perform six code summarization approaches on them to explore the impact of comment categories on code summarization.  They reported that no models perform the best for ``why'' and ``property'' comments among the six categories.
%Gro et al. \cite{DBLP:conf/kbse/GrosSDY20} reported that a naive information retrieval approach can meet or exceed reported performance from neural models. 
Most of the previous work focused on assessing the model effectiveness in terms of error types, comment intentions, preprocessing operations, and dataset size, while our work aims to investigate the model effectiveness on difficult levels of the code summarization task, complementing the existing studies. In addition, we report {\RQone} and {\RQtwo} in the code-comment dataset, which could provide a sounder foundation for existing work. 

\section{Conclusion}
\label{sec:conclusion}

We propose a taxonomy of data preprocessing noises in four popularly used benchmark datasets for code summarization, which contains 12 different types
of noise. We further build a rule-based cleaning tool for detecting noisy data of each category.
Experiments show that, the tool can accurately detect noises in our manually annotated data. 
We then apply the cleaning tool to the four benchmark datasets, and assess their data quality. The results show that noisy data extensively exist in the four widely-used benchmark datasets (ranging from 31\% to 66\%).
Finally, we investigate the impacts of noisy data on three types of code summarization models (\textit{i.e.}, NNGen, NCS, and Rencos) by comparing their performance trained with datasets before and after the cleaning. The results show that the performance of three existing models trained with the filtered benchmark datasets improves BLEU-4 by 27\%, 21\%, and 24\%, ROUGE by 19\%, 11\%, and 16\%, METEOR by 19\%, 7\%, and 16\%, CIDEr by 46\%, 19\%, and 33\%, respectively. 
We release our tool as a python library, named CAT, to facilitate relevant research in both academia and industry.

In our future work, we plan to extend our research methodology to other text generation tasks in software engineering such as commit message generation and code synthesis.  
\section*{ACKNOWLEDGMENTS}
We sincerely appreciate anonymous reviewers for their constructive and insightful suggestions for improving this manuscript. 
This work is supported by the National Key Research and Development Program of China under Grant No. 2018YFB1403400, the National Science Foundation of China under Grant No. 61802374, 62002348, 62072442, 614220920020 and Youth Innovation Promotion Association Chinese Academy of Sciences.

\vspace{0.8cm}

\balance

\bibliographystyle{ACM-Reference-Format}
\bibliography{ref}
\end{document}